\documentstyle[graphics,graphicx,times,subfigure]{mn2e}
\input{psfig}

\newcommand {\gsim}{\mbox{$\:\stackrel{>}{_{\sim}}\:$} }
\newcommand {\lsim}{\mbox{$\:\stackrel{<}{_{\sim}}\:$} }

\def\ion#1#2{#1{\sc#2}}
\def\HI{\ion{H}{i}}
\def\loss{lines of sight}
\def\citet#1{}
\def\lya{Ly-$\alpha$}
\def\etal{et al.}
\def\ls{L_{\rm s}}
\def\dlos{\langle d_{\rm LOS}\rangle}
\def\nlos{N_{\rm LOS}}

\def\Sec#1{Section~\ref{sec:#1}}
\def\Fig#1{Figure~\ref{fig:#1}}
\def\Eq#1{Equation~(\ref{eq:#1})}
\def\Tab#1{Table~\ref{tab:#1}}

\begin{document}

  \title[The topology of IGM at high redshift]{Recovering the  topology of the
IGM at z$\sim$2}

  %\titlerunning{}%% give here short title %%
  \author[Sara Caucci \etal]{S. Caucci$^{1}$, S. Colombi$^{1}$, 
    C. Pichon$^{1,2}$, E. Rollinde$^{1}$, P. Petitjean$^{1}$, 
    T. Sousbie$^{1,2}$\\
    ${}^1$ Institut d'Astrophysique de Paris \& UPMC,
    98 bis boulevard Arago, 75014 Paris, France\\
    ${}^2$ Centre de Recherche Astrophysique de Lyon, 
    9 avenue Charles Andr´e, 69561 Saint Genis Laval, France\\
  }
  %\authorrunning{Sara Caucci \etal}
  %\offprints{S. Caucci}
  %\institute{Institut d'Astrophysique de Paris - 98bis bd Arago,
  %75014 Paris, FRANCE}

  \maketitle

  \begin{abstract}
    
    We investigate how well the 3D density field of neutral hydrogen in the 
    Intergalactic Medium (IGM)
    can be reconstructed using the Lyman-$\alpha$ absorptions observed
    along lines of sight to quasars separated by arcmin distances
    in projection on the sky. 
    We use cosmological  hydrodynamical  simulations 
    to compare the topologies of different fields: dark matter, gas and 
    neutral hydrogen optical depth and to investigate how well the topology of
    the IGM can be recovered from the Wiener interpolation method 
    implemented by Pichon et al. (2001).
    The {\it global} statistical and topological properties of  the
    recovered field are analyzed quantitatively through the
    power-spectrum, the probability distribution function (PDF),  the Euler
    characteristics, its associated critical point counts 
    and the filling factor of underdense regions. The {\it local} geometrical
    properties of the field are analysed using the local skeleton by
    defining the concept of inter-skeleton distance.

    As a consequence  of the nearly lognormal nature of the 
     density distribution at the scales under consideration,  
    the tomography is best carried out on
   the  logarithm of the density rather than the density itself. 
    At scales larger than  $\sim 1.4 \dlos$, where $\dlos$ is the mean
   separation between lines of sight, 
   the reconstruction accurately recovers  the topological 
   features of the large scale density distribution of the gas,
   in particular the filamentary structures: the
   inter-skeleton distance between the reconstruction and the exact
   solution is smaller than $\dlos$. 
   At scales larger than the intrinsic smoothing length 
   of the inversion procedure, the power spectrum 
   of the recovered H{\sc i} density
   field matches well that of the original one and the low order
   moments of the PDF are well recovered as well as the shape of the Euler
   characteristic.
   The integral errors on the PDF  and the critical point counts 
   are indeed small, less than 20\% for a mean line of sight 
   separation smaller than $\sim$2.5 arcmin.  
   The small deviations between the reconstruction
   and the exact solution mainly reflect departures from 
   the log-normal behaviour that are ascribed to highly 
   non-linear objects in overdense regions.
   \end{abstract}

  \begin{keywords}
    methods: statistical,  hydrodynamical simulations -- 
    cosmology: large-scale structures of universe,
    intergalactic medium -- quasars: absorption lines
  \end{keywords}

\section{Introduction}

  The  structure and composition of the intergalactic medium (IGM) has
  long been studied using the \lya\ forest in QSO absorption spectra
  \cite{rauch98}. The progress made in  high resolution 
  Echelle-spectrographs has led to a consistent
  picture in which the absorption features are related to the
  distribution of neutral  hydrogen through the Lyman transition lines
  of \HI.  
  Hydrogen in the IGM is highly ionized (Gunn \& Peterson,
  1965). Its photoionization equilibrium in the expanding IGM
  establishes a tight correlation between neutral and total hydrogen
  density and numerical simulations have confirmed the existence  of this
  correlation. They  have also shown that the gas density traces the
  fluctuations of the DM density on scales larger than the Jeans
  length (see for example Cen \etal\ 1994, Petitjean \etal\ 1995,
  Miralda-Escud\'e \etal\ 1996, Theuns \etal\ 1998, Viel, Haehnelt \&
  Springel 2004).

  As we will show in the first part of this work,  the statistical and
  topological properties of the IGM and of the dark matter
  distributions are the same, so that  recovering the
  three-dimensional distribution and inferring the topological
  properties of the IGM allows us to constrain the properties of the
  dark matter distribution as well.

  Although topological tools have been introduced only relatively
  recently in cosmological analysis, they have been  used extensively
  to characterize the topology of large scales structures as revealed
  by the three-dimensional distribution of galaxies in the local
  universe (see for exemple Gott \etal\ (1986), Vogeley \etal\ (1994),
  Protogeros \& Weinberg (1997), Trac \etal\ (2002), Park \etal (2005)
  and Sousbie \etal\ (2006) for the topological analysis of galaxy
  surveys).  The outcome of such an analysis is a {\it quantitative}
  description of the complex appearance of the distribution of the
  matter in the universe, with its network of clump, voids, filaments
  and sheet-like structures.  The study of the topology using 
  galaxy surveys is attractive because of their large volume and  the
  huge number of objects they contain.
  However the clustering of highly non linear  objects (galaxies,
  clusters of galaxies or QSOs) is biased compared to the underlying
  clustering of dark matter fluctuations that we wish to constrain
  (see Kaiser 1984).  This biasing results from a
  complicated and delicate competition between a variety of processes
  which are often too complicated to be tractable analytically.
   Besides, the maximum redshift in surveys is low (in the
  analysis of the SDSS data made by Park \etal\ 2005, the maximum redshift
  is $z=0.1654$), so that this kind of analysis can be done only in
  the local Universe, where the fluctuations have already entered the
  highly non-linear regime.

  Given the strong correlation existing between  dark matter
  distribution and the low-density intergalactic medium, one could
  probe the underlying distribution of  matter  via
  the  signature produced by  diffuse hydrogen  in quasar spectra,
  namely  absorption features observed in the \lya\ forest.
   Indeed,  absorption spectra provide a  picture complementary to those
  drawn by galaxy surveys to infer the large scale distribution of the
  matter in the Universe, since the absorption features produced by
  the IGM in \lya\ forest can be detected also at large redshift and
  since  the IGM probes the low density range, whereas the galaxy 
  distribution  does not.  Eventually, higher density   contrasts can
  be recovered from the analysis of the \lya\ forest if higher order
  transitions are included in the analysis;  for example, the
  Ly-$\beta$ transitions should allow us to probe density contrasts up
  to $\delta\approx 15$.

  The flux along a single line-of-sight towards a quasar only
  provides  one dimensional information, which can be used to
  constrain the fluctuation amplitude and the matter density (Nusser
  \& Haehnelt 1999, Rollinde \etal\ 2001, Zaroubi \etal\ 2006). The
  transverse information, found in pairs of quasars, has been used to
  study the extension of the absorbing regions (e.g. Petitjean \etal\
  1998; Crotts \& Fang 1998; Young, Impey \& Foltz 2001; Aracil \etal\
  2002) and the geometry of the Universe at $z\sim 2$ (Hui, Stebbins
  \& Burles 1999; McDonald \& Miralda-Escud\'e 1999; Rollinde \etal\
  2003, Coppolani \etal\ 2006).

  Given a set of lines of sight (LOSs) toward a group of QSOs with
  small angular separation, inversion methods can be used to recover
  the three-dimensional distribution of  low density gas, as
  demonstrated in  Pichon \etal\ (2001).  They showed that the visual
  characterization of the density field (with its network of
  filaments, clumps, voids and pancakes)   is correctly reproduced if
  the mean separation  between the LOSs is less than  $\dlos\leq 5$
  Mpc.

  In this paper we test  quantitatively whether such an inversion
  can recover the global properties of connectivity of the density
  field,  using topological tools such as the Euler characteristic and
  the probability distribution function.

  The paper is  organized as follows.  In \Sec{definition},  the Euler
  characteristic is defined as an alternate critical  points count and
  implemented for a Gaussian field. The difference  between the
  topological properties of the dark matter, of the gas  and of the
  observed optical depth is then discussed using  outputs of a
  hydrodynamical simulation (\Sec{euler}) and relying on different
  statistical tools.  In \Sec{inversion}, the ability to reconstruct
  the global topology of the three dimensional distribution from a
  simple  Wiener interpolation of a discrete group  of \loss\ is
  considered.   Finally, \Sec{conclusion} summarizes the
  results of this paper and discusses some possible improvements
  of the method as well as observational constrains from future
  surveys.

  %===========================================================
 \section{The Euler characteristic:  an alternate critical point count}
\label{sec:definition}
  %===========================================================

  This paper makes use of various statistical tools, namely   the 
  PDF, the Euler characteristic, the skeleton  and  related
  estimators such as the first cumulants of  the PDF (connected
  moments), critical point counts and the filling factor,  to
  characterize  the topology of the large scale density
  distribution.   These tools will also
  be used to test the efficiency of 
  reconstructing the density field from a grid of QSO sight-lines and
  in particular
  the ability to reproduce the connectivity  of the large scale
  structures.

  Following Colombi, Pogosyan \& Souradeep (2000, hereafter CPS),
  this section  introduces the Euler characteristic, ${\tilde \chi}^ {+}$,
  as an alternate critical point count in an overdense excursion with
  density contrasts larger than a threshold $\delta_{\rm TH}$.  It is
  shown how the behavior of ${\tilde \chi}^{+}$ is related  to 
  connectivity in the field.  The numerical implementation used to
  measure it is described and tested on Gaussian random realizations.

  %===========================================================
  \subsection{Definition of the Euler characteristic} \label{sec:euldef}
  %===========================================================

  Let $\delta({\mathbf x})$ be a scalar function 
  defined in a 3D volume $V$.  Given a threshold value
  $\delta_{\rm TH}$,  consider the excursion set $E^+$  formed by the
  points ${\mathbf x}$ with $\delta({\mathbf x})\ge\delta_{\rm TH}$,
  as expressed by the following equation:
  \begin{equation}\label{eq:excursion_set}
    E^+\equiv\{\mathbf x | \delta({\mathbf x})>\delta_{\rm TH}\}\,.
  \end{equation}
  The analysis of the geometrical properties of  points that belong
  to  the excursion set $E^+$ as a function of $\delta_{\rm TH}$ gives
  information about the global topology of the scalar field
  $\delta({\mathbf x})$ and allows for the  characterization  of large
  scale structures.

  A simple qualitative link can be  established  between the
  distribution of critical points (defined by  
  $\nabla\delta=\mathbf{0}$), a   on the one hand, and
   connectivity on the other,   which are 
   related to {\it local}  and {\it global} 
   properties  of the excursion set respectively.
  If one considers over-dense regions,
  connectivity   happens  along ridges (filaments) passing through
  saddle    points and connecting local maxima.  The same reasoning
  can be applied to  under-dense regions where minima are connected
  through tunnels (pancakes) via another kind of saddle point.  This
  idea is in fact supported on  rigorous grounds by  Morse theory
  (see Milnor 1963). The Morse theorem establishes the link between
  the distribution of critical points and the global connectivity of
  the excursion set, via the Euler characteristic.  This quantity
  represents the integral of the  Gaussian curvature over an
  iso-density surface  that marks the boundary of the excursion set
  (see for exemple Gott, Melott \& Dickinson, 1986). It
  is usually defined
  as the following  count (see for example Mecke, Buchert \&
  Wagner, 1994, for details):
  \begin{equation}
    {\tilde \chi}^{+}={\rm connected\ components}-{\rm tunnels}+{\rm
    cavities}.
  \end{equation}
  According to Morse theorem, it can also be expressed as a linear
  combination of the number of critical points of different types that
  are found in the excursion set as a function of $\delta_{\rm TH}$.

  To be more specific, let us consider the critical points of the
  field.  For these points, the Hessian matrix, whose  components are
  given by:
  \begin{equation}\label{eq:hessian}
    {\mathcal H}_{i,j} = \frac{\partial^2\delta}{\partial x_i\partial
    x_j}\,,
  \end{equation}
  is calculated and its eigenvalues are estimated.  According to the
  number  of negative eigenvalues, $I$, of the Hessian matrix, the local
  structures of the field can  be classified in  the following way: a
  clump, a filament, a pancake and a void corresponding to 
  $I=$3, 2, 1 and 0 respectively. The Morse theorem states that the Euler
  characteristic can be expressed as a count of the number
  of critical points belonging to each of these four classes:
  \begin{equation}\label{eq:euler}
    {\tilde \chi}^{+} = N_{I=3}-N_{I=2}+N_{I=1}-N_{I=0}\,,
    \label{eq:chidef}
  \end{equation}
  where $N_{I=i}$ is the number of critical points with $i$ negative
  eigenvalues.
  With this approach, it is sufficient to determine  the number
  distribution of the four kinds of critical point. However
  this  differential method requires the field under consideration to
  be sufficiently smooth and non degenerate.  To this end, in 
  the subsequent analyses of this paper,  the field will be smoothed with
  a  Gaussian window (using standard FFT technique),
  \begin{equation}
     W(r)=\frac{1}{(2\pi)^{3/2} L_s^3} \exp\left(
     -\frac{r^2}{2\ls^2}\right),
     \label{eq:smoothingwindow}
  \end{equation}
  of sufficiently large size $L_s$ compared to the sampling grid pixel
  size in order to minimize the impact of numerical artefacts coming
  from the discretization of the field on a grid (see, e.g.,  CPS for a
  thorough analysis of measurement issues). In what follows, the
  smoothing scale used to measure the Euler characteristic  is always
  larger than $0.01\times N_{\rm pix}$ grid pixels where $N_{\rm
  pix}=256$ pixels is the box resolution of  the simulations.  In
  principle this smoothing scale is large enough to have an unbiased
  measurement of the Euler characteristic.  The prescription of CPS is
  used to detect and classify  critical points.  This method
  involves locally fitting a second order hypersurface on the smoothed
  density field, while  taking into account  each point on the grid
  under consideration and its 26 neighbors.

  %===========================================================
\subsection{Interpretation of the Euler characteristics}
  \label{sec:interpr_euler}
  %===========================================================

  \begin{figure}
    \centerline{\psfig{figure=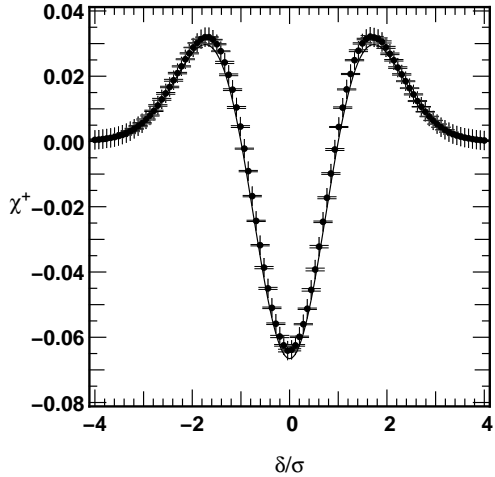,width=6.7cm}}
    \caption[]{Mean Euler characteristics, $\chi^+$ (see
      Eqs.~\ref{eq:euler} and \ref{eq:normeuler}),  for a Gaussian
      random field (GRF, points with small errorbars) as a function of
      the density threshold, $\delta/\sigma$, compared to the
      theoretical prediction (Doroshkevich, 1970, smooth curve).  The
      mean  is carried over 5 realizations of a GRF whose power
      spectrum is given by a power-law with spectral index $n=-1$ on a
      $256^3$ grid, while additional smoothing is performed with a
      Gaussian window of size  $5$ pixels.  }
    \label{fig:euler_gaussian}
  \end{figure}
  \begin{figure}
    \centerline{\psfig{width={6.3cm},figure=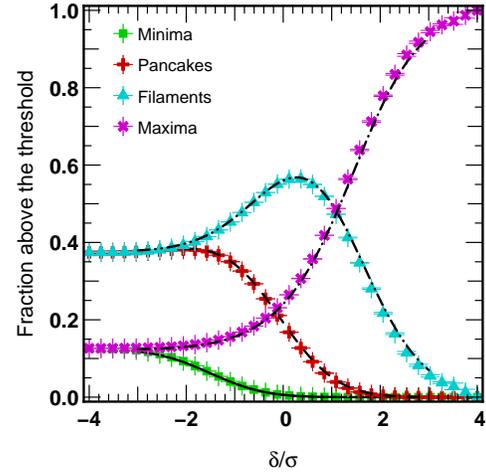}}
    \caption{Evolution of the number of critical points entering the
      computation of the Euler characteristic for the GRF considered
      in Fig.~\ref{fig:euler_gaussian}.  The fraction of different
      types of critical points above the threshold is plotted as a
      function of $\delta/\sigma$ and each distribution is compared to
      the analytical prediction.  Again, symbols with error bars
      represent the mean over 5 realizations of the same GRF, while
      the smooth curves give the analytical prediction (which can be
      easily derived from Bardeen et al. 1986). }
    \label{fig:criticalpts_gaussian}
  \end{figure}

  For clarity, let us recall here the interpretation of the shape  of
  the Euler characteristic as a function of density threshold (CPS).
  Let us first study   the simple  case of a Gaussian random field
  (GRF).  The analytic predictions for a GRF are given, for example,
  in Doroshkevich (1970) (see also, Schmalzing \& Buchert 1997).  In
  what follows,  a slightly different normalization from
  Eq.~(\ref{eq:chidef})  is used:  the volume independent quantity
  \begin{equation}
    {\chi}^{+}={\tilde \chi}^{+}/N_{\rm tot},
    \label{eq:normeuler}
  \end{equation}
  where $N_{\rm tot}=\sum_i N_{I=i}$ is the total critical point count
  in the volume considered, in the limit $\delta_{\rm TH} \rightarrow
  -\infty$.

  In \Fig{euler_gaussian},  the numerical estimates of  the Euler
  characteristic are given as a function of the density  threshold
  $\delta_{\rm TH}=\delta/\sigma$ where
  $\delta=(\rho-\bar{\rho})/\bar{\rho}$ is the  density contrast  and
  $\sigma={\rm rms}(\delta)$ from five Gaussian random field (GRF)
  realizations (points with error bars)  whose power spectrum is given
  by a power-law with spectral index $n=-1$, i.e. $P(k)\propto k^n$.
  The result is compared with the analytic prediction (solid line).
  The shape of the curve as a function of  the density threshold can
  be  understood from \Eq{euler} and 
  \Fig{criticalpts_gaussian} which displays the  critical point
  counts.  At very low values of the threshold ($\delta/\sigma \la
  -4$)  the excursion set includes almost all points and,  due to the
  symmetry between high and low-density regions,  the number of minima
  (pancakes) compensates the number  of maxima (filaments) so that the
  Euler characteristic approaches zero.  When  the value of the
  threshold is increased, local minima first drop out  of $E^+$,
  creating cavities and thus increasing the  value of $\chi^+$.  At
  $\delta/\sigma\gsim -2$, pancakes start to drop out too and  cavities
  connect together, thus the value of  $\chi^+$ decreases,  reaching
  its minimum at $\delta/\sigma\approx 0$.   In the range
  $0\lsim\delta/\sigma\lsim2$ filaments also drop out, breaking up the
  ridges to create isolated clusters,  thus increasing $\chi^+$ again.
  Finally in the region $\delta/\sigma\gsim 2$ only clumps are  found
  to lie in the excursion set, but they are progressively lost as the
  threshold increases,  explaining  the final decrease of the
  curve.

  This simple analysis shows how the features seen in the  Euler
  characteristic are closely related to the network of   filaments and
  pancakes that connect clumps and voids.

  %===========================================================
\section{From dark matter to optical depth}\label{sec:euler}
  %===========================================================

  The Lyman-$\alpha$ absorption lines observed in QSOs spectra
  and produced by the \HI\ structures intercepted by the
  line-of-sight  can be used to study the topology of the Universe
  at high redshift ($z\gsim 1-2$). However, the information derived  from
  observations of QSOs spectra is  more directly related to the  \HI\
  optical depth, whereas  here the aim is  to
  constrain the underlying dark matter density field for which
  theory makes direct predictions.
  Hence, one has to rely on simulations in order to calibrate the
  relation between the density field  of the neutral 
  hydrogen and that of the   dark matter.

  In this  section  we first present the hydrodynamical simulations
  used in the present work and we then analyse the shape of the PDF and the
  Euler characteristic of the three density fields (dark matter,
  gas and \HI) and of the optical depth field, explaining
  how these curves are related.

  %===========================================================
\subsection{Numerical simulations}\label{sec:simul}
  %===========================================================

 We analyse a cosmological hydrodynamical simulation that evolves both dark
  matter particles and a gaseous component  to
  study  the global topology of the intergalactic medium at redshift $z=2$.  
  The
  dynamical evolution and the physical properties of the gas and the
  of the \HI\ component are calculated  taking into account the heating and
  cooling processes and the effect of the ionizing UV background in a
  standard way. The corresponding Particle-Mesh (PM) code used to
  perform the simulation is described in detail in Coppolani \etal\
  (2006).

  In this run, the standard $\Lambda$CDM model is assumed with a set of
  cosmological parameters given by: $\Omega_{\rm m}=0.3$ and
  $\Omega_{\rm \Lambda}=0.7$, while the assumed baryon density is
  $\Omega_{\rm b}=0.04$.  The Hubble constant is H$_0=70\,{\rm
  km\,s^{-1}\,Mpc^{-1}}$ and the amplitude of the fluctuations of the
  matter density field in a sphere of radius $8\ h^{-1}$ Mpc is
  $\sigma_8=1$. While the other cosmological parameters are roughly in
  agreement with recent observational constraints, the value of
  $\sigma_8$ is somewhat large compared to the value suggested by WMAP
  (see Spergel \etal, 2007). However, this should not have any
  incidence on the results derived in this paper.

  The simulation involved $512^3$ dark matter particles in a
  box with periodic boundary conditions of comoving size
  $L_{\rm box}=40$ Mpc. The gaseous component was
  also followed on a $512^3$  grid which was used  to compute
  gravitational forces.  Although this simulation marginally resolves
  the Jeans length of the gas, Coppolani \etal\ (2006) checked with
  higher resolution runs that numerical convergence was achieved at
  small scales.

   Although $512^3$  grid points were available, this resolution was
  degraded to a $256^3$ resolution (using standard donner cell
  procedure), in order to make the calculations more  tractable.
  Obviously this additional smoothing makes the effects of
  subclustering within the Jeans length irrelevant.  Therefore the
  gaseous component should  be nearly  indistinguishable from the dark
  matter component.

  The main limit in these analyses remains the box size, which is
  still small and only allows for a  fair statistical measure at
  scales $\ls$ larger than $L_{\rm max}\sim L_{\rm box}/10$, i.e. 4
  Mpc. Indeed finite volume effects are known to become significant for
  $\ls \ga L_{\rm max}$ for standard statistics such as the
  probability function (see, e.g., Colombi, Bouchet \& Schaeffer 1994)
  and the Euler number (see, e.g., CPS). For the reconstruction, the
  typical separation $\dlos$ between lines of sight defines a natural
  smoothing scale $\ls \simeq \dlos$.  Note that, unfortunately, 
  the upper bound of $\ls \sim 4$ Mpc  corresponds to a lower bound on
  $\dlos$ in current state of the art observations \cite{rollinde03},
  but one can expect to lower this limit in future  surveys
  \cite{TheunsSrianand06}.  Hence the following analyses are performed 
  in the range $2\ {\rm Mpc} \leq \ls \leq 4$ Mpc.

 \begin{figure*}
    \begin{center}
      \hspace{0.4cm}
      \psfig{width=10cm,figure=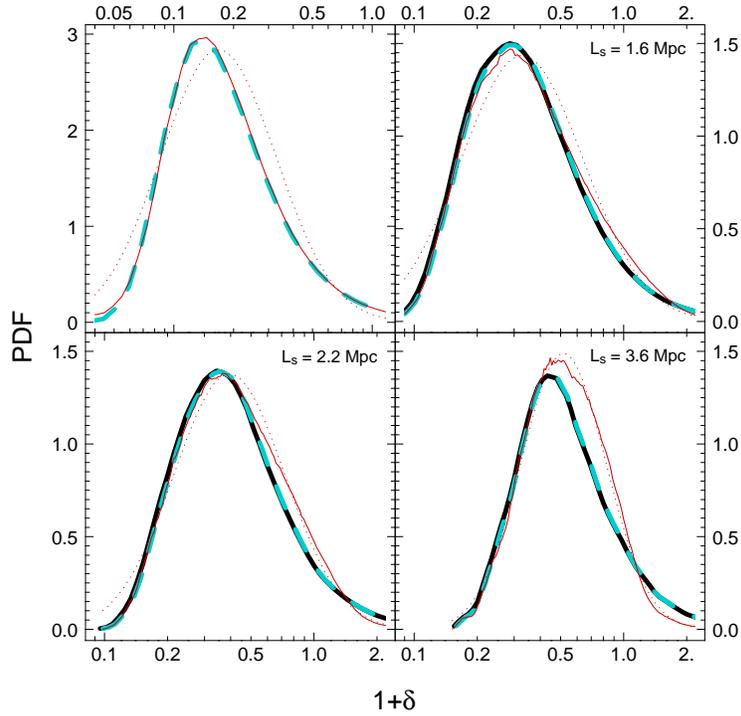}
    \end{center}
    \caption{Probability distribution function  of density fields
      at different smoothing scales ({\it from left to right, top to
      bottom}, no smoothing, $\ls$=1.6, $2.2$ and $3.6$ Mpc). The {\it
      thick solid, thick  dashed} and {\it thin solid} curves
      correspond to dark matter, gas and \HI\ (rescaled according to
      Eq.~\ref{eq:powerlaw}), respectively.  The dotted curve is a
      best fit of  a lognormal distribution  to the thin solid curve,
      showing that  all these PDFs are reasonably close to lognormal,
      a property that will  be  useful  for  the reconstruction.
      In the unsmoothed case, the gas and H~{\sc i} PDFs match very well 
      for $1+\delta\lsim1$ but depart from each other at higher density.
      The apparent very good match in the unsmoothed case comes from
      the fact that the  un-shocked part of
      the intergalactic medium totally dominates the part of the PDF
      which is visible in this panel. The match between \HI\ and gas
      PDFs decreases with increasing smoothing scale, due 
      to ``mixing effects'', as explained in  the main text.  
      Note finally that the dark matter
      is not displayed in the unsmoothed panel  because the result
      would be contaminated by the cloud-in-cell  interpolation used
      to compute the density on the grid.  }
    \label{fig:PDFs_densfields}
 \end{figure*}
 \begin{figure*}
   \begin{center}
     \hspace{0.2cm}
     \psfig{width=15cm,figure=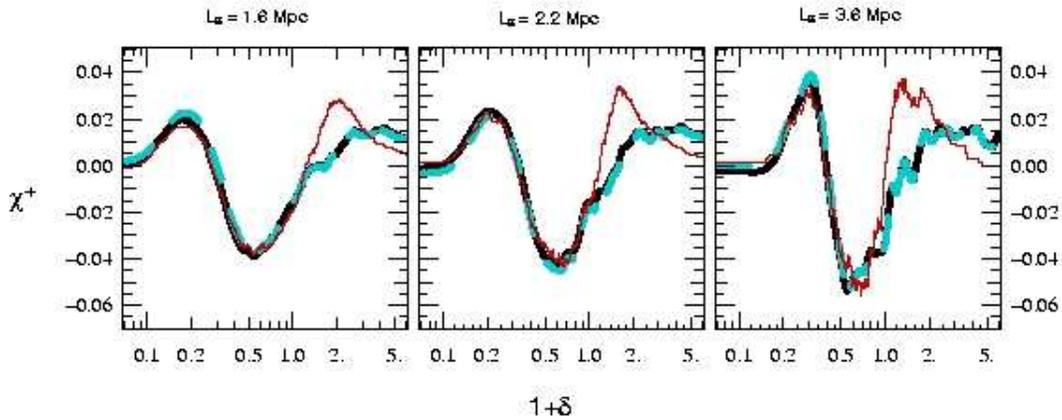}
   \end{center}
   \caption{Same as in Fig.~\ref{fig:PDFs_densfields} but for the
     measured Euler characteristic as a function of density threshold
     at different smoothing scales. Again,  the {\it thick solid,
     thick dashed}  and {\it thin solid} curves  correspond
     respectively to dark matter, gas and  \HI\ (rescaled according to
     Eq.~\ref{eq:powerlaw} with the values  $(A,\alpha)$ given in
     Table~\ref{tab:bestfit}). While the curves for the dark matter
     and for the gas superpose exactly at all smoothing scales and all
     values of the threshold, the \HI,  even after the scaling is
     applied, behaves in a different way in the high density region.
     As explained in the text, this is a consequence of the presence
     of shocks and condensed objects, whose effect is a change in
     connectivity properties.}
   \label{fig:Euler_densfields}
 \end{figure*}

  %===========================================================
\subsection{PDF and
    Euler characteristic of physical density fields}
  %===========================================================

  It this Section we compare the large-scale distribution and
  the topological properties
  of the different density fields (dark matter, total amount
  of gas and neutral gas) by looking at their probability
  distribution function (PDF) and their Euler characteristic
  ($\chi^+$).  Our knowledge of the physics of the intergalactic
  medium is used to perform the analysis and to link the 
  distributions of \HI\ and  H. Indeed,  the observations  give 
  access  to the \HI\ optical depth through absorption spectra.
  We also consider  thermal  broadening and
  redshift distortion effects.

  \subsubsection{From dark matter to \HI: IGM equation of state}
  It is well known that on scales larger than the Jeans length
  the distribution of the gas 
   follows the distribution of dark matter, so that their statistical and
  topological properties  are expected to be the same at these scales.
  This is checked by comparing the PDF and the Euler
  characteristic of the two density fields smoothed using  different
  values of $\ls$, as shown in
  Figs.~\ref{fig:PDFs_densfields} and \ref{fig:Euler_densfields}.
  Note  the  agreement of the PDFs and of the Euler characteristics  
  of the two fields for all  values of 
  the smoothing scale considered,  a result which can be expected 
  since the scaling regime probed is largely above
  the Jeans length of the gas.

  \begin{figure*}
    \centering
    \subfigure[Gas density and temperature]
	      {\includegraphics[width=11.cm]{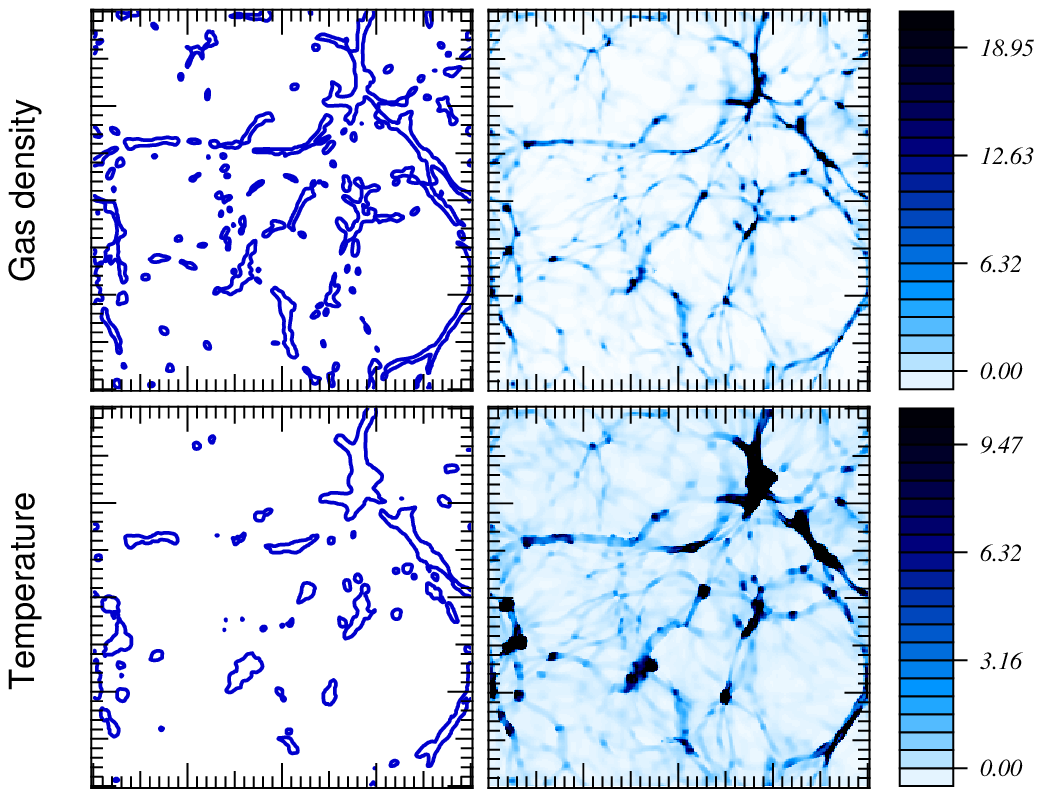}
      \label{fig:shocks_a}}
    \subfigure[Ratio R=${\tilde\rho_{\rm HI}}/\rho_{\rm gas}$]
	      {\includegraphics[width=11.cm]{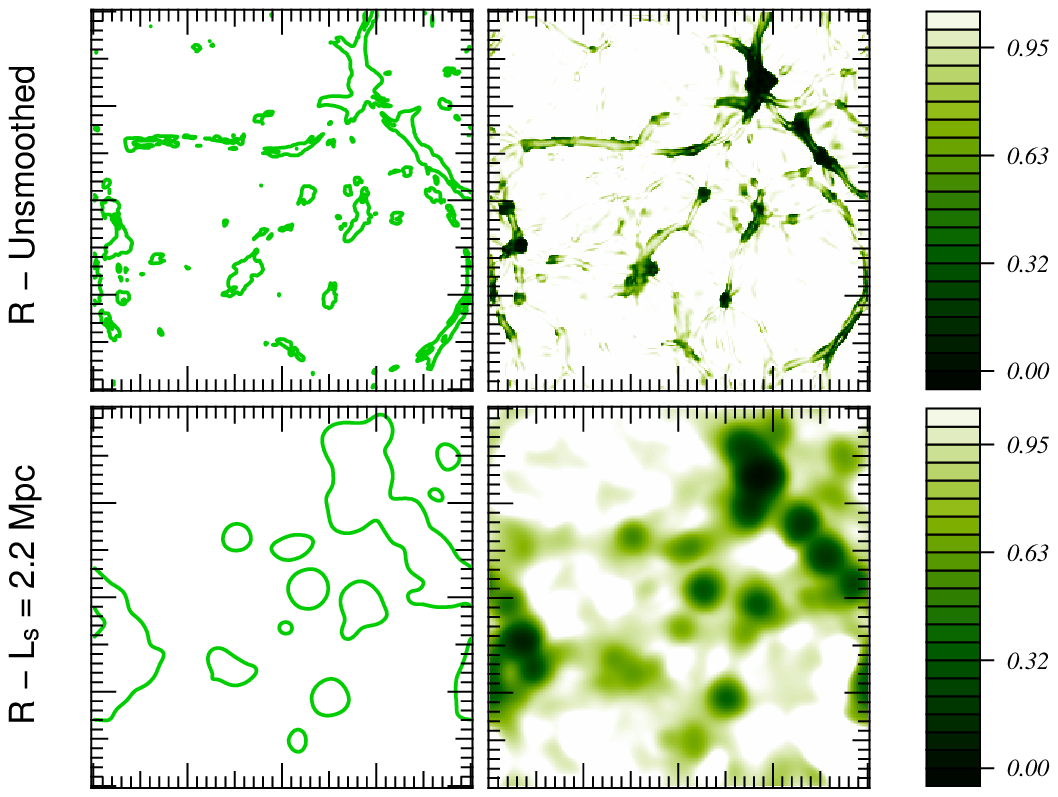}
      \label{fig:shocks_b}}
    \caption{{\it Top}: Gas density and temperature (in units  of
      $10^4$ K) spatial distributions in a one pixel ($\approx 15.6$
      kpc)  slice. The intensity of the fields is color-coded with the
      scale given on the right. The panels on the left give the
      contours corresponding to $\delta=1$ for the density and to
      $T/10^4=2$ for the temperature. {\it Bottom}: For the same slice
      as above we show on the right  the spatial distribution of the
      ratio $R={\tilde\rho_{\rm HI}}/\rho_{\rm gas}$ for the
      unsmoothed field ({\it up}) and for the field  smoothed with a Gaussian
      window of size (FWHM)  $\ls=2.2$ Mpc ({\it down}).  The  color scale  is
      such   that darker regions correspond  to low values of $R$. On
      the left the contours correspond to $R=0.7$.}
      \label{fig:shocks_tot}
  \end{figure*}
  \begin{figure*}
    \begin{center}
      \hspace{0.45cm}
      \psfig{width=11.5cm,figure=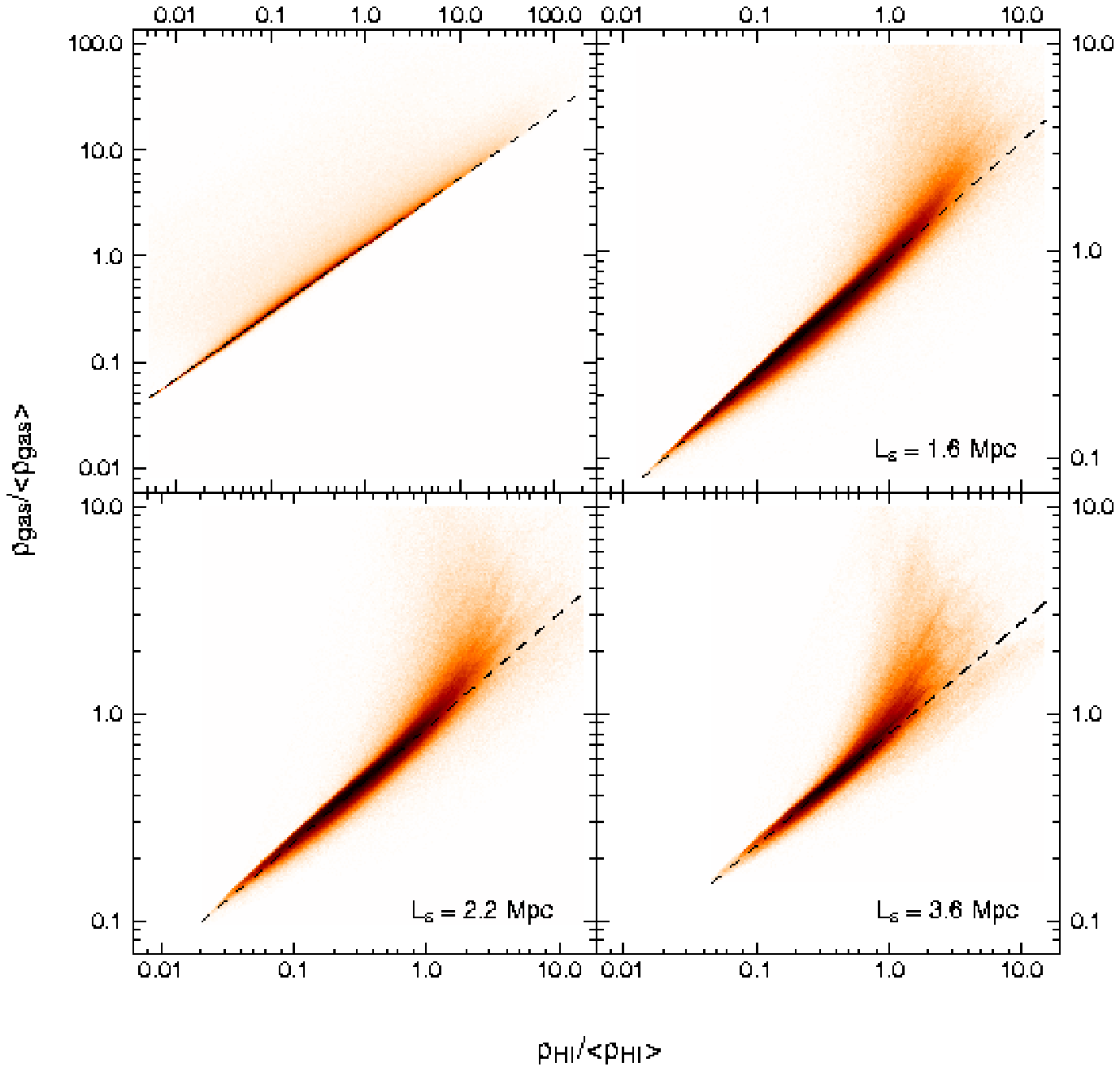}
    \end{center}
    \caption{Scatter plots displaying the relation between the
      gas density and the \HI\ density at different smoothing scales.
    The dashed black lines in each panel represent the best fit,
    following Eq.~(\ref{eq:powerlaw}), with the parameters $(A, \alpha)$
    given in \Tab{bestfit}. Note that the dispersion increases when
    the smoothing scales increases, due to the mixing effect discussed
    in \Fig{shocks_tot}. 
    }
    \label{fig:scatter}
  \end{figure*}

  The comparison of  the distribution of the neutral gas (\HI) with
  that of the total amount of gas and  the dark matter calls for a
  slightly more elaborate approach, given the non-linearity involved
  in the expression that relates the distribution of the gas to the
  distribution of \HI.  In fact, numerical simulations  support
  the idea that a tight correlation exists between neutral and total
  hydrogen density (Cen \etal\ 1994, Miralda-Escud\'e \etal\ 1996,
  Theuns \etal\ 1998, Viel, Haehnelt \& Springel 2004).  This
  correlation is expected to follow a power-law of the form
  \begin{equation}\label{eq:powerlaw}
    \rho_{\rm gas}\approx A\cdot (\rho_{\HI })^\alpha\,.
  \end{equation}
   We thus introduce here a new density field $\tilde\rho_{\rm \HI}$ 
  defined as the right-hand-side of Equation~(\ref{eq:powerlaw}), 
  so that $\tilde\rho_{\rm \HI}\equiv A\cdot (\rho_{\HI })^\alpha$.
  In what follows this new density field will be used in order to
  approximate the density of the gas. However, 
  Equation~(\ref{eq:powerlaw}) is not fulfilled in the whole range
  of $\rho_{\HI }$ values.
  
   To illustrate this,  Fig.~\ref{fig:shocks_a} displays
  the gas density distribution (top), 
  with its network of filaments outlined
  in the left panel with a contour corresponding to $\delta=1$,
  and the temperature distribution in units of $10^4$ K (bottom)
  for which we have drawn  the contour   corresponding to 
  $(T/10^4)=2$ in the left panel. Note that along filaments 
  and at their intersection the
  gas is  hot. This indicates that shock waves propagate
  along filaments, rising the temperature and ionizing the
  gas. This is confirmed by Fig.~\ref{fig:shocks_b} which shows the ratio
  $R={\tilde\rho_{\rm HI}}/\rho_{\rm gas}$  measured directly in the
  $256^3$ grid (top), and after a Gaussian smoothing with
  a window whose size is $\ls=2.2$ Mpc (bottom). In both cases, the
  panels on the left show the contours relative to $R=0.7$.
  To complete the picture, let us consider Fig.~\ref{fig:scatter}
  which shows the scatter between $\rho_{\rm gas}$ and ${\rho}_{\rm \HI}$
  for different smoothing scales, as indicated in each panel.
  As expected, the tightness of the correlation is very high in
  underdense and moderately dense regions, but shock heating on the one
  hand and the formation of condensed objects on the other
  produce a  significant scatter (where $R<1$) 
  along densest filaments and at their intersection (in
  clusters).  For the purpose  of the reconstruction,  some smoothing
  is required.  Unfortunately, smoothing also mixes these regions  with
  the  un-shocked part of the intergalactic medium.  This is 
  confirmed, in a qualitative way, by the slices shown in 
  Fig.~\ref {fig:shocks_b}.   More quantitatively, for the
   fields $R$ shown in \Fig{shocks_b} we have 
  calculated the fraction of the volume occupied by the regions with 
  $R<0.7$ (i.e. the volume of the  regions enclosed by the
  contours in the left panels). For the slice shown, 
  this fraction is $f(R<0.7)=0.07$ and $f(R<0.7)=0.19$ 
  for the un-smoothed and the smoothed case respectively,
  while when the whole three dimensional boxes are considered, 
  the fraction of volume occupied by shocked regions 
  are $f(R<0.7)=0.05$ and $f(R<0.7)=0.11$.
  As a result of such mixing, the tightness of the
  correlation is weakened, but  remains good as shown in
  Fig.~\ref{fig:scatter}.  However, the best fit values of the 
  parameters $A$ and $\alpha$ changes slightly when the field
  is smoothed (see Table~\ref{tab:bestfit}). 
  We fit these values with the low density tail of the 
  PDF (see \Fig{PDFs_densfields}). 
  As expected, the higher   density tail match worsens with smoothing.

  \begin{table}
    \begin{center}
      \begin{tabular}{|c|c|c|}
    \hline
    $\ls$ & $A$ & $\alpha$\\
    \hline  \hline
    Unsmoothed  & 1.275  & 0.63 \\
    1.6 Mpc     & 0.915  & 0.56879 \\
    2.2 Mpc     & 0.85   & 0.55209\\
    3.6 Mpc     & 0.795  & 0.5389 \\
    \hline
      \end{tabular}
    \end{center}
    \caption{Values of the parameters A and $\alpha$ entering in
    the scaling relation between gas and \HI\ (see Eq.~\ref {eq:powerlaw})
    as a function of the smoothing scale $\ls$.}
    \label{tab:bestfit}
  \end{table}

  Given that  the scaling relation (\ref{eq:powerlaw}) is monotonous, 
  should it  apply exactly, the topology of the neutral gas 
  should be  exactly the same as of the total gas/matter 
  distribution. However, given the dispersion of this relation,  
  one expects the Euler characteristic of the $\tilde\rho_{\rm \HI}$ 
  field  to depart from that of $\rho_{\rm gas}$ for large density 
  contrasts. This is confirmed by Fig.~\ref{fig:Euler_densfields}:
  a nearly perfect agreement is found between the gas
  and \HI\ for $\delta \la 0$, while differences become significant
  at larger values of the density contrast.
  Increasing the smoothing length (i.e. going from
  left to right  in Fig.~\ref{fig:Euler_densfields})
  worsens the match, as expected, but this
  is in part lost in the noise due to finite volume effects.
  Note that $\chi^+$ measured in \HI\ is, in the $\delta > 0$
  regime, more peaked than for the total gas.
  This agrees with intuition, since galaxies form in filaments:
  in  these highly condensed objects, gas concentrates
  and cools down. Hence the \HI\ density becomes significant
  again inside these clumps, but is depleted in
  their surroundings due to shock heating  as can be seen from
  Fig.~\ref{fig:shocks_tot}. The resulting
  distribution of \HI\ in filaments is therefore
  expected to be more clumpy than the total gas i.e. less
  efficiently connected,
  resulting in a larger increase of $\chi^{+}$ for $\delta > 0$.
  The estimates made inside filaments are however certainly not free
  of numerical artefacts since they are limited by the simulation's  spatial
  resolution (following accurately the formation
  of condensed objects requires much higher spatial resolution than
  our simulation). Therefore although one  can definitely trust
  the $\delta\la 0$ measurements, the results derived for
  $\delta > 0$ are likely to yield the right qualitative behavior,
  but are certainly quantitatively biased.

  In the next Sections,  the $\delta >0$ disagreement will be ignored
  and it  will be assumed  that the scaling relation (\ref {eq:powerlaw})
  is always valid, keeping in mind  the limitation of such
  an assumption. Hence, reconstructions will be performed 
  on the optical depth without attempting to directly recover  the gas
  distribution.

  \begin{figure*}
    \begin{center}
      \hspace{0.4cm}
      \psfig{width=9.5cm,figure=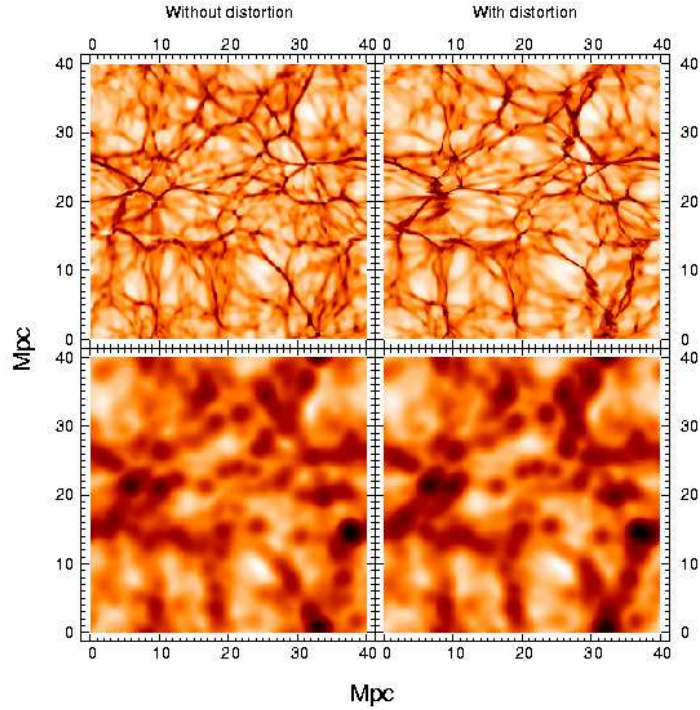}
    \end{center}
    \caption{The effect of redshift distortion on the HI density.  The
      same slice (whose width is 6 pixels, corresponding to 0.94 Mpc)
      of the \HI\ density contrast is shown without distortion ({\it
      left panels}) and with distortion ({\it right panels},  using
      the infinitely distant observer approximation with a distortion
      along the  $x$ axis) in the case where the fields are not
      smoothed ({\it top panels}) and when the fields are smoothed at
      $\ls=1.6$ Mpc (bottom panels).  }
    \label{fig:slice_effect_reddist}
  \end{figure*}
  \begin{figure*}
    \begin{center}
      \hspace{0.2cm}
      \psfig{width=15cm,figure=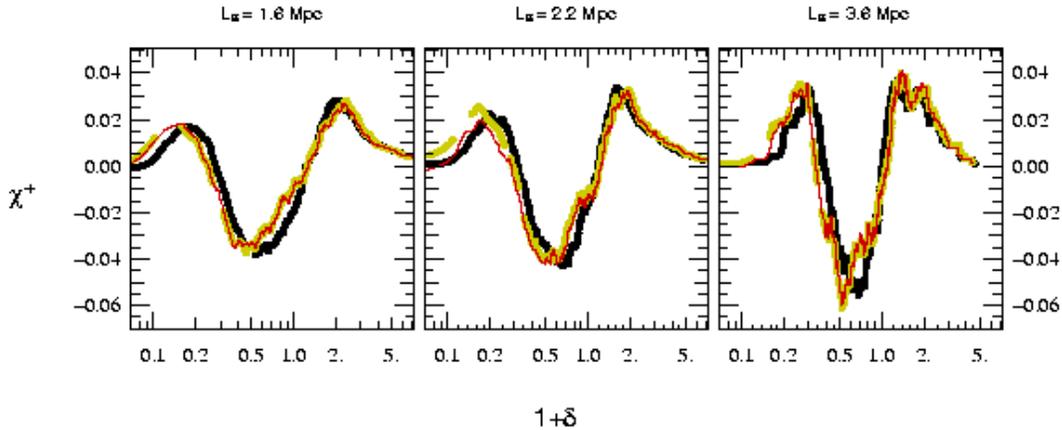}
    \end{center}
    \caption{Effect of redshift distortion on the Euler
      characteristic of \HI\ at different smoothing scales:
      $\ls=1.6$ Mpc, 2.2 Mpc and 3.6 Mpc. The solid black
      line is for \HI\ without any distortions, the dashed
      yellow line has been obtained by including only the effect of
      peculiar velocities, while for the red thin line both
      redshift distortion and thermal broadening are taken into
      account. 
    }
    \label{fig:euler_effect_reddist}
  \end{figure*}

\subsubsection{From \HI\ to optical depth: redshift space distortions  and
  thermal broadening}

  In the above discussion we argued that the main features of  dark
  matter topology, as traced through the Euler characteristic and the
  probability density function, can be recovered through the topology
  of the \HI\  for small density contrasts, $\delta \la 0$.  However,
  along a line-of-sight, the optical depth is in fact observed in
  redshift space, where distortions induced by the peculiar motions
  operate. Moreover, the profiles of absorption lines are broadened at
  small scales by the effect of the temperature.  Since the thermal
  broadening is important only at scales of the same order or smaller
  than the Jeans length,  this second effect should be negligeable in
  the scaling range considered in this paper, since it will be swept 
  out by the smoothing.  On the contrary,  redshift distortion should
  {\it a priori} not be neglected.

  In theory, it is possible  to  partially correct   for redshift
  distortion effects  (see for instance PVRCP).  However, the
  corresponding treatment of the peculiar velocities involves a
  simultaneous deconvolution of the \HI\ density field with the
  velocity field on top of  the inversion discussed in next section.
  This requires not only a prior for the density field but also for
  its correlation with the  peculiar velocity field and makes  the
  inversion quite convolved and this would go beyond the scope of this
  paper.  In what follows, it is shown that in fact redshift
  distortions have a small effect on the topology of the overall
  density distribution for the probed scales; they shall thus be
  neglected in the reconstruction part of this work. Moreover, one of
  the interesting outcomes of the reconstruction is to predict  the
  positions of filaments in the three dimensional matter distribution.
  Cross-correlation of such a distribution with for instance the
  observed distribution of galaxies at high redshift can in fact be
  also performed in redshift space.

  Figure \ref{fig:slice_effect_reddist} displays the \HI\ distribution
  in real and redshift space with and without smoothing: the main
  effect of redshift distortion on \HI\ is an enhancement of large
  scale density contrasts orthogonally to the line of sight  due to
  large scale motions (this is the so-called Kaiser effect,
  e.g. Kaiser 1987): the ``voids'' (underdense regions) are more
  pronounced, and the filaments orthogonal to the LOS are more
  contrasted. There is as well a small scale ``finger of god'' effect,
  due to internal motions inside large dark matter haloes, but it is
  not very pronounced at such a high redshift, and is in amplitude of
  the same order of thermal broadening.  Note however, that non
  trivial shell crossings can still occur, e.g. two filaments crossing
  each other thanks to peculiar  velocities, but this effect remains
  small, and is clearly damped out by  smoothing; after  smoothing
  only the  Kaiser effect remains.

  These qualitative arguments are illustrated by
  \Fig{euler_effect_reddist}. The measured Euler characteristics
  before and  after redshift distortion differ  only slightly.  When
  the redshift distortion is taken into account, for $\delta \la 0$,
  a shift towards the left is induced (dashed curve) as compared  to
  the non-distorted  case (solid one); while the opposite occurs for
  $\delta \ga 0$ (although in the latter case, the effect seems to be
  nearly insignificant).  This shift remains quite small as argued
  before. Note as well that thermal broadening (thin curve) is totally
  negligeable.

  Finally, one last point should be mentioned.  When one considers
  real absorption spectra, instrumental noise has to be taken into
  account in the analysis. This noise, combined with saturation of the
  flux of the \lya\ absorption lines arising in high density regions
  (with $\delta\ga 10$)  can complicate the interpretation of the
  measurements.  In this case, some of the information about the
  intensity of the density   field cannot be recovered, unless, say,
  Lyman-$\beta$ is also available.  In this work, however, the main
  interest lays  in reproducing the low-density part of the \HI\
  distribution, for which the relation  (\ref{eq:powerlaw}) holds and
  for which the topology of the underlying dark matter distribution is
  theoretically constrained.  In this regime, the \lya\ lines are not
  saturated, thus a complete treatment of saturation effects in high
  density regions is not required for the aim of the present work.

%======================================================================
\section{Topological and statistical properties of the recovered fields}
  \label{sec:inversion}
%======================================================================

  The absorption spectrum  towards a quasar gives  access to
   one-dimensional information i.e. the
  optical depth along the line of sight (LOS)
  towards the QSO. However, if a set of LOSs towards
  a group of quasars is available, the information along each LOS  can be
  interpolated  to construct
  a 3-dimensional optical depth field.

  In this Section we first briefly outline  the inversion technique
   implemented to recover the optical depth
   and describes how to set the parameters that enter the
   inversion procedure.  We then check how the reconstruction performs
   by measuring various statistical quantities, in particular the
   PDF of the density field and its
   Euler characteristic.  As argued in the previous Section, the focus
   is on the optical depth: no attempt is made to recover
   the gas or dark matter distribution directly.  Thermal
   broadening, redshift distortion and effects of saturation or
   instrumental noise are neglected. Given these assumptions,
   studying the optical depth distribution is then equivalent to
   studying the \HI\ density distribution, $\rho_{\HI}$.

%======================================================================
\subsection{The inversion method: Wiener interpolation}
%======================================================================
The technique used to interpolate the optical depth
  field  between \loss\ is described and discussed in
  details in PVRCP.

  Let {\bf D} be a 1-dimensional array representing the data set (i.e.
    the values of $\gamma_{\rm LOS}=\ln(\rho_{\HI})$
  along the LOSs, which we  assume  to be parallel to each
  other); we call {\bf M} the 3-dimensional array of the parameters
  that need  to be estimated (here the values of $\gamma_{\rm
  3D}=\ln(\rho_{\HI})$ in the 3-dimensional volume) by  fitting the
  data.  Wiener interpolation reads (see Eq.~20 of PVRCP), assuming
  the noise is uniform and uncorrelated,
  \begin{equation}
    {\bf M}={\bf C_{\rm MD}}\cdot({\bf C_{\rm DD}}+{\bf N})^{-1}
    \cdot {\bf D}\,,
    \label{eq:inversioncombo}
  \end{equation}
  where ${\bf N}=n^2 {\bf I}$ is the diagonal noise contribution,
  ${\bf C_{\rm MD}}$ is the mixed parameters-data covariance matrix
  and ${\bf C_{\rm DD}}$ is the data covariance matrix:
  \begin{equation}
    {\bf C_{\rm MD}}={\bf C_{\gamma_{\rm 3D}\gamma_{\rm LOS}}}\,,
    \quad {\bf C_{\rm DD}}=
	 {\bf C_{\gamma_{\rm LOS}\gamma_{\rm LOS}}}\,.
  \end{equation}
  Here  an {\it ad-hoc} prior is used and a Gaussian shape for the
  covariances is assumed. In this cases the  matrices 
  ${\bf C_{\gamma_{\rm 3D}\gamma_{\rm LOS}}}$ and 
  ${\bf C_{\gamma_{\rm LOS}\gamma_{\rm LOS}}}$ are given by
  \begin{eqnarray}
    {\bf C}(x_1,x_2,{\bf x_{1\perp}},{\bf x_{2\perp}}) =
    \sigma^2\times\exp\Big(-\frac{(x_1-x_2)^2}{L_x^2}\Big)\times
    \nonumber\\ 
    \exp\Big(-\frac{|{\bf x_{1\perp}}-
      {\bf x_{2\perp}}|^2}{L_T^2} \Big)\,,
    \label{eq:correl_matrix}
  \end{eqnarray}
  where $(x_i,{\bf x}_{i\perp})$ represents the coordinates of the
  points  along and perpendicular to the LOSs respectively, $L_x$ and
  $L_T$ are correlation lengths along and perpendicular to the LOSs,
  while $\sigma^2$ quantifies the typical {\it a priori}  fluctuations in a
  volume of size $L_x\times L_T^2$.  The  meaning and choice of
  these parameters will be discussed further in \S~\ref {sec:paraminv}.

  Note that the shape of the covariance matrix can be calculated 
  with a  more sophisticated approach. This would
  involve the use of theoretical priors relying on our knowledge of
  large scale structure dynamics. If, for instance, the reconstruction
  was performed on the pure dark matter density  contrast, one could
  be tempted to derive these correlation matrices from the non-linear
  power spectrum obtained, for example, by Peacock \& Dodds (1996),
  given a cosmological model.  Here, a    simpler interpolation scheme
  is used. This  scheme  has  the advantage of depending only on
  three tuning parameters: the assumed typical overall signal-to-noise
  ratio, $\sigma/n$, and two typical lengths in the interpolation,
  $L_x$ and $L_T$.

  %===================================================================
\subsection{Choice of the parameters in the interpolation}
  \label{sec:paraminv}
  %===================================================================

  Each reconstruction is performed on a number $\nlos$ of LOSs
  extracted at random  from the simulation box. Since the distant
  observer approximation is implemented, all the LOSs  are parallel.
  For a given value of $\nlos$, the mean inter-LOS distance, $\dlos$,
  reads:
  \begin{equation}\label{eq:d_los}
    \dlos\equiv\sqrt{L_{\rm box}^2/\nlos}.
  \end{equation}
  This parameter obviously defines a natural scale in the
  reconstruction: one cannot, intuitively, expect to reconstruct
  details of the  distribution at scales $\la \dlos$, at least
  in the direction orthogonal to the LOSs.

  The meaning of the parameters $L_T$ and $L_x$ in
  Eq.~(\ref{eq:correl_matrix}) is then quite straightforward.  The
  correlation lengths $L=L_T$ and $L=L_x$ stabilize the inversion by
  ensuring the smoothness of the reconstruction. In order to avoid the
  formation of fictitious structures, the transverse correlation
  length must be of the order of the mean separation between the LOSs,
  $L_{\rm T} \sim \dlos$ (we have chosen to take it exactly equal to $\dlos$),
  while the choice of the longitudinal correlation length depends on
  the problem considered. Since redshift distortion is
  not a concern in the present work, this parameter can be  chosen to be of
  the order of the Jeans length in order to avoid  
  information loss for small scales along the LOSs, here $L_x=0.4$ Mpc.

  \begin{figure*}
    \centerline{Raw fields}
    \vskip 0.2cm
    \centerline{\psfig{height={7.4cm},figure=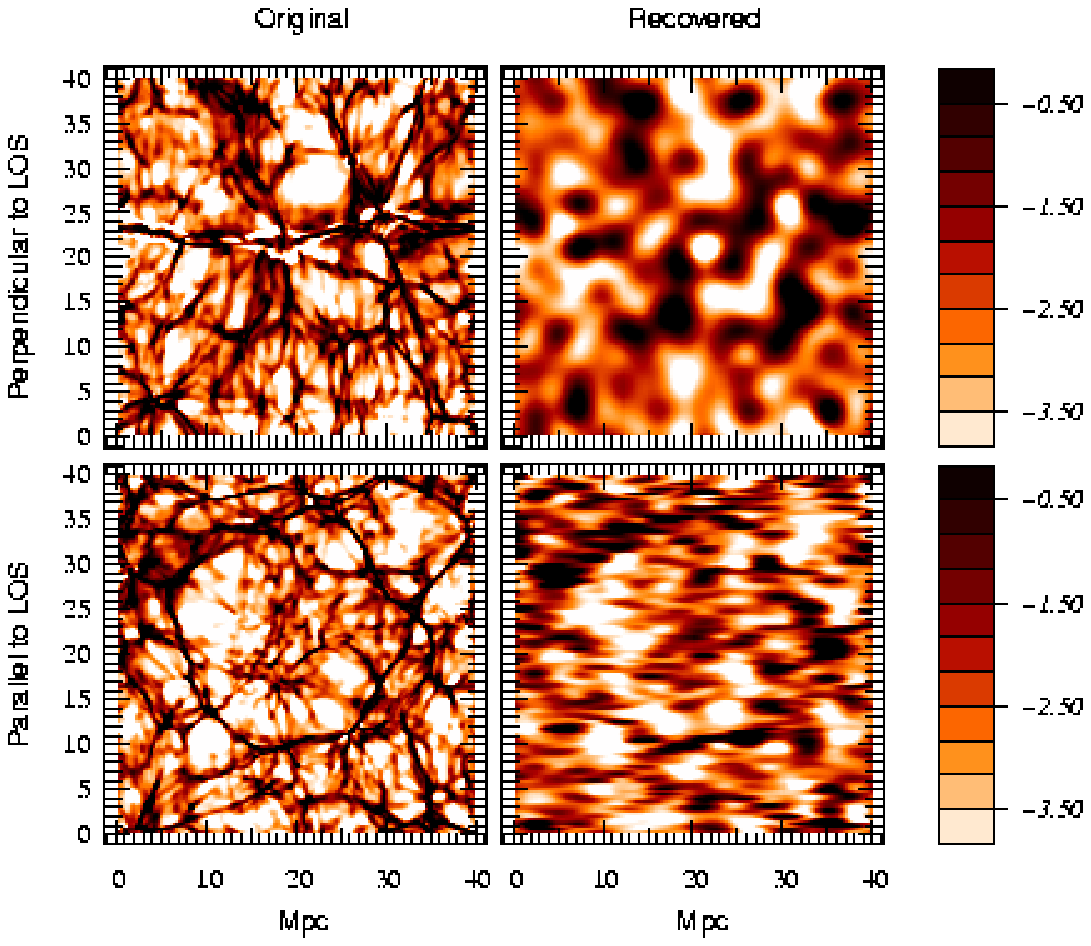}}
    \hspace{0.2cm}
    \centerline{Smoothed fields in logarithmic space \hskip 5cm
    Smoothed fields in linear space}
    \vskip 0.2cm
    \centerline{\psfig{height={7.4cm},figure=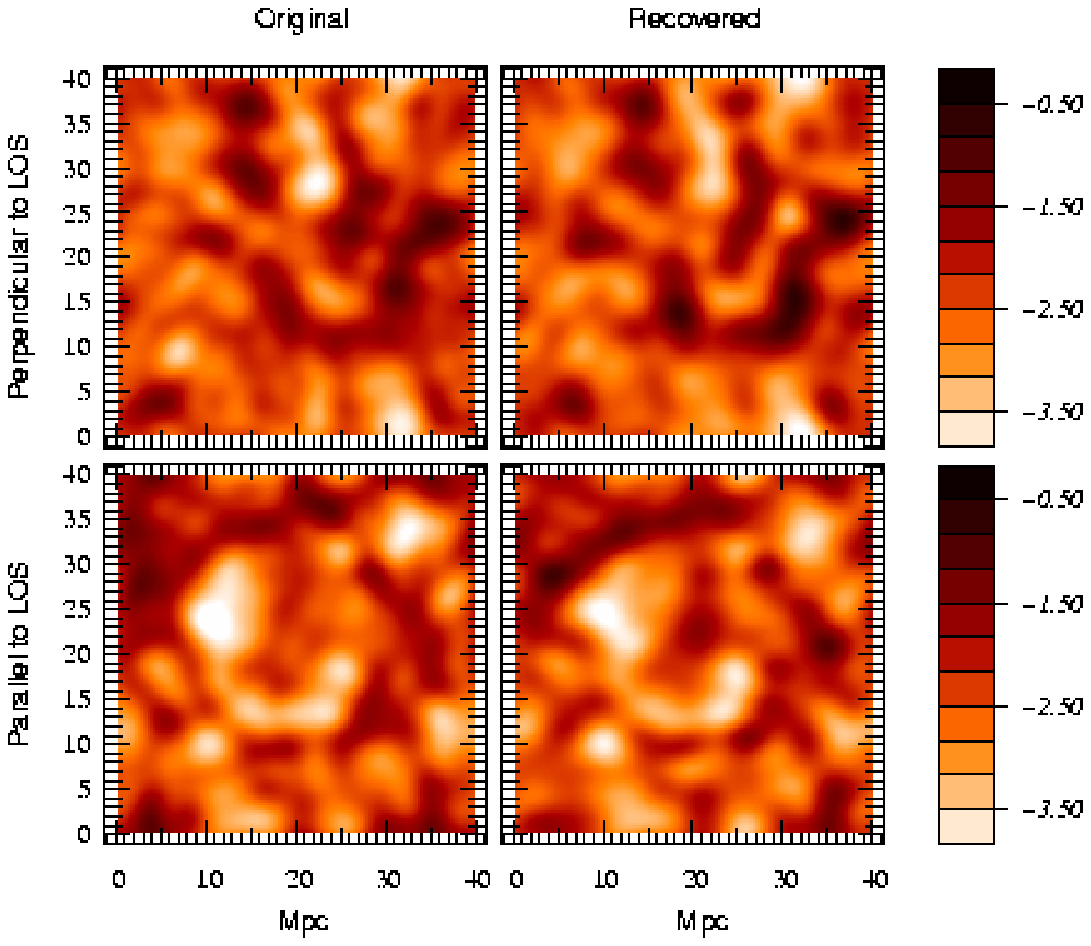}
      \psfig{height={7.4cm},figure=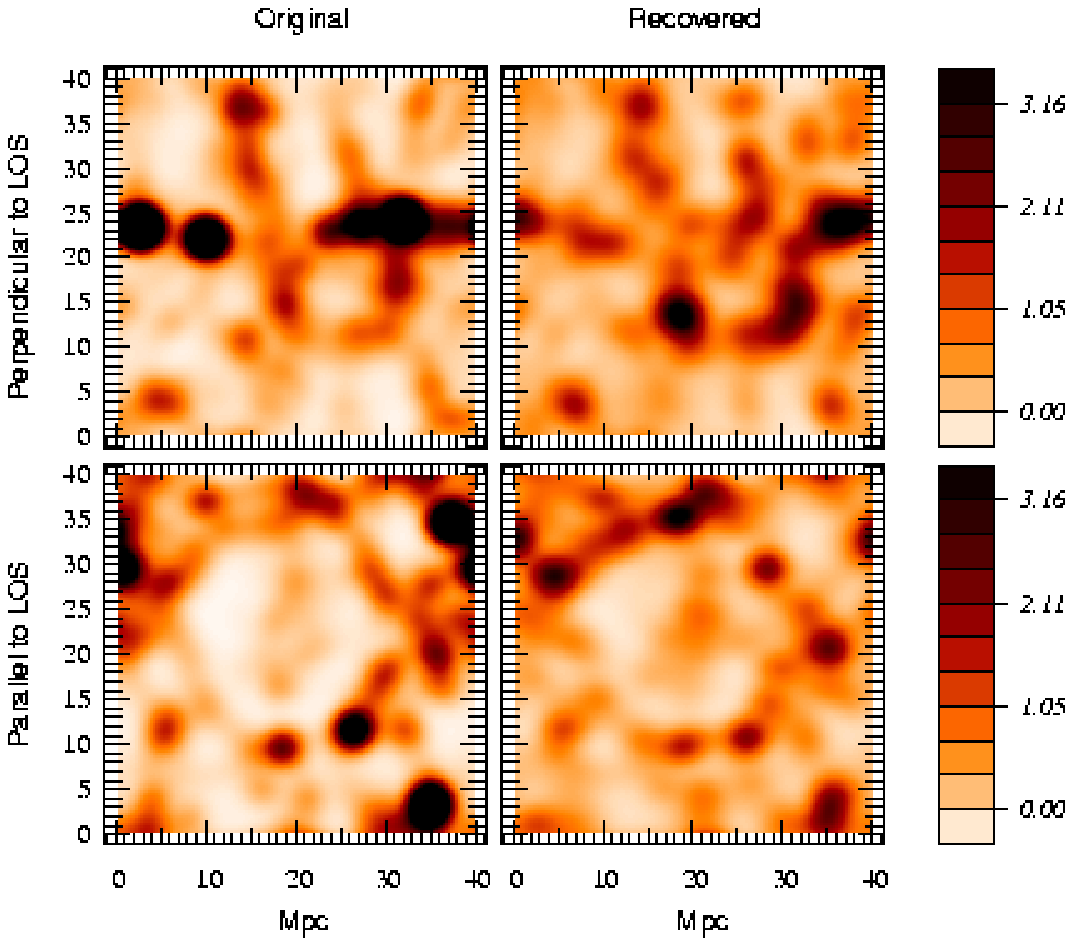}}
    \caption{Qualitative comparison between the original \HI\ density
      field in terms of $\gamma=\ln(1+\delta)$ (left column in each
      group of 4 panels) and the recovered one (right column in each
      group of 4 panels)  in a thin slice  (the thickness of the
      slice is 8 pixels, corresponding to 1.25 Mpc).  Higher
      densities correspond to darker colors. The recovered field has
      been obtained by inverting a set of $N_{\rm LOSs}=320$ random
      LOSs (mean separation $\langle d_{\rm LOS}\rangle=2.24$ Mpc) 
      taken through the original (unsmoothed)  density 
      field. In each group of panels, the  first row
      corresponds to a slice orthogonal to the LOSs, while the second
      row  corresponds to a slice parallel to the LOSs.  {\it Upper
      group:} the raw $\gamma$ fields for the original box and for
      the reconstruction.  {\it Lower left group:} same as the upper
      group but after smoothing (in the logarithmic space) 
      with a Gaussian window of radius
      $L_{\rm s}=\sqrt{2}\langle d_{\rm LOS} \rangle = 3.17$ Mpc.  
      {\it Lower right group:}  same as the lower left one, 
      but smoothing is now applied directly to the density field,
      $1+\delta=\exp(\gamma)$, instead of its logarithm and the
      normalization is slightly different  (see Eq.~\ref{eq:normalexp}
      of Appendix~\ref{appendix}).  }
    \label{fig:slice_recorig}
  \end{figure*}

From a practical point of view,
  the variance parameter $\sigma$ of the correlation matrix
   fixes the relative contribution of signal to noise in
  Eq.~(\ref{eq:inversioncombo}), $\sigma/n$. In our reconstruction,  
  only ideal   LOSs are considered. Thus,  strictly speaking,  
  there is no instrumental noise or saturation effects.  
  However, the inversion of the matrix  ${\bf
  C_{\rm DD}}+{\bf N}$ is  numerically unstable when ${\bf N}$
  is set to zero, given the finite sampling and  the degeneracy of the
  matrix, Eq~(\ref{eq:correl_matrix}), close to its diagonal, $(x_1,{\bf
  x_{1\perp}}) \simeq (x_2,{\bf x_{2\perp}})$.  In practice one has to
  ``tune'' the signal-to-noise ratio, $\sigma/n$, to obtain the best
  compromise between numerical stability and ``exactness'' of the
  final reconstruction.  This choice is {\it ad-hoc}:  $(\sigma/n)^2$
  is the estimated variance  $\sigma^2(L_T,L_x)$ of the underlying
  field in a box of size $L_T\times L_T\times L_x$.  This is
  equivalent to assuming that, as the noise goes to zero,  the inverse of
  the  non-reduced second order correlation (in the appropriate
  units) is used,  ${\bf I}+{\bf C}_{\rm DD}$, instead of the reduced
  one, ${\bf C}_{\rm DD}^{-1}$, to perform a  stable reconstruction.

  In this work we estimate directly  $\sigma^2(L_T,L_x)$   from the
  simulation.  It is however important to note that
  $\sigma^2(L_T,L_x)$ can in principle be derived from the LOSs alone
  by measuring the 1-D power-spectrum of $\rho_{\HI}$. From
  this 1-D power-spectrum, one can indeed infer a 3-D power-spectrum
  with  standard deconvolution methods and then an estimate of
  $\sigma^2(L_T,L_x)$ by the appropriate integral on the 3-D
  power-spectrum.

  The measured values of $\sigma(L_T,L_x)$ are listed in 
  Table~\ref{tab:set_param}.  They are of the order unity: the assumed
  signal-to-noise ratio is about  one in the regime considered 
  here. Hence, in practice,  the {\it ad-hoc} procedure used  to perform the
  inversion does not  change significantly by
  including  the contribution of the actual  instrumental
  noise.  However, in this case the presence of the saturated regions 
  in the Lyman-$ \alpha$ spectra remains a problem.

  \begin{table}
   \begin{center}
      \begin{tabular}{|c|c|c|c|}
    \hline
    $\nlos$ & Separation  & $L_T$ & $\sigma$\\
            & (arcmin)    & (Mpc)\\
    \hline  \hline
    400 & 1.33  &  2     & 1.12\\
    320 & 1.49  &  2.24  & 1.17\\
    225 & 1.77  &  2.67  & 1.23\\
    200 & 1.88  &  2.83  & 1.25\\
    145 & 2.2   &  3.32  & 1.29\\
    120 & 2.42  &  3.65  & 1.31\\
    100 & 2.65  &  4     & 1.34\\
    \hline
      \end{tabular}
   \end{center}
   \caption{Parameters used in the reconstructions performed in this
     paper. The longitudinal correlation length as been fixed to
   the value $L_x=0.4$ Mpc for all the reconstructions.}
   \label{tab:set_param}
  \end{table}

  Finally, note that due to the large size of the  matrices, 
reconstructions can only be performed by partioning the simulation box in 
  blocks of  smaller size, that contain $N_{\rm sub}^3$ grid points
  with $N_{\rm sub}=32$. The reconstruction is performed
  on each block individually.
  In order to avoid edge effects, neighboring   patches are overlapped
  by adding a buffer region in which LOSs still contribute. In
  this way, the {\it a priori} correlation ensure continuity 
  between adjacent patches.  The size of the buffer region
  is chosen to be $n_{\rm over} \simeq 2 L_T$  (in grid pixel units),
  which implies a typical residual contamination of edge effects due
  to the partitioning of less than 2 percent.

 % ========================================================================
 \subsection{Bias in the reconstruction}
 \label{sec:biasrec}
 % ========================================================================

  \begin{figure*}
    \centerline{\psfig{width={15cm},figure=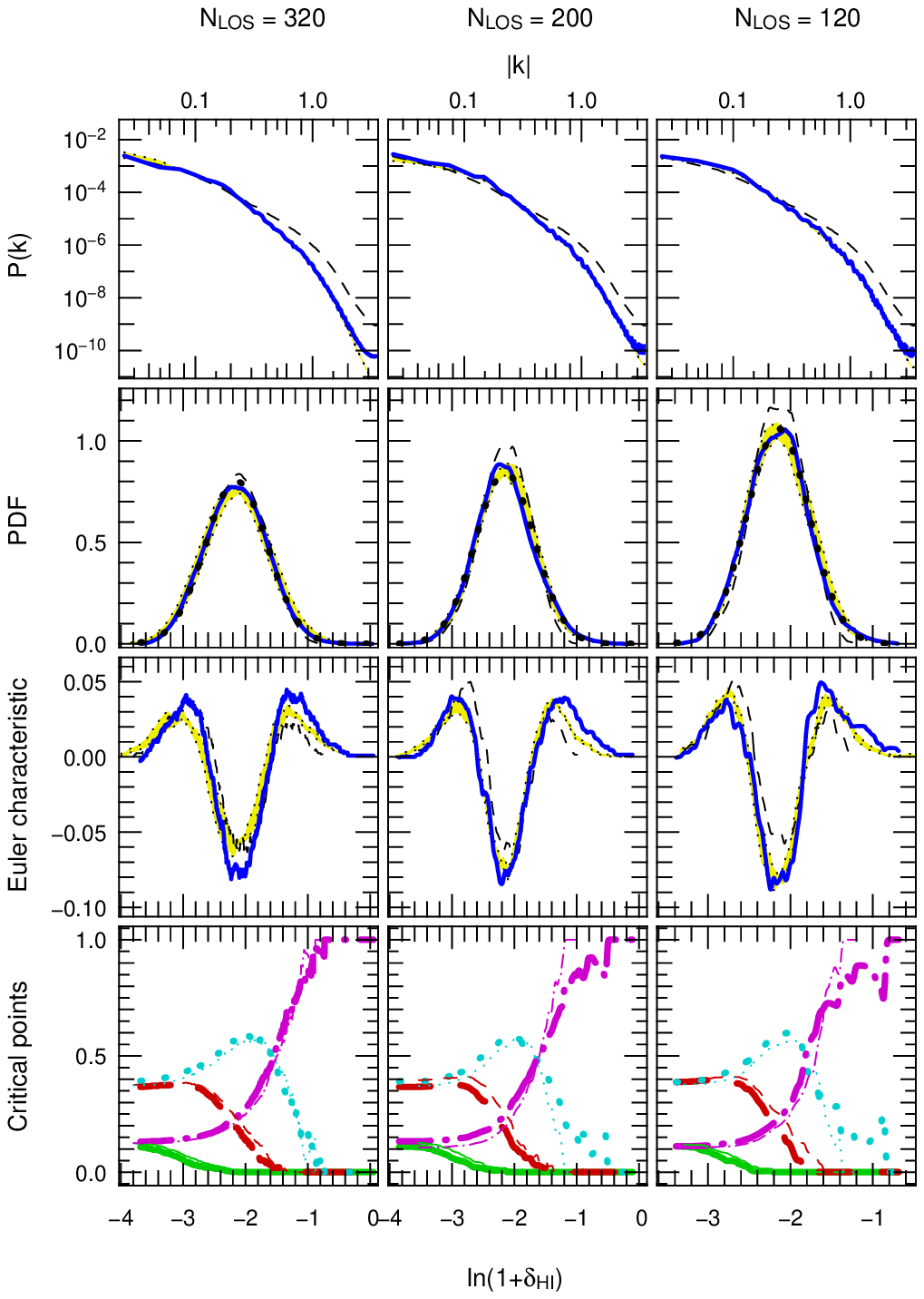}}
    \vspace{-0.5cm}
    \caption{Statistics and topology in logarithmic space, i.e.  in
      terms of $\gamma_{\HI}=\ln(1+\delta_{\HI})$, for three
      different reconstructions performed with $\nlos=320$, $\nlos=200$
      and  $\nlos=120$, from left to right respectively.  {\it First
      row of panels:}  the power-spectrum of $\gamma_{\HI}$ as a
      function of wavenumber indicated at the top.  In each panel, the
      {\it thin dashed} curve represents the power spectrum of the
      original field while the {\it thick solid} line is the power
      spectrum of the (unsmoothed) recovered field. The light shaded 
      region corresponds to the scatter between five realizations of
      Gaussian fields (GRFs) with the same power  spectrum as the
      reconstruction.  The wavenumber $k$ is expressed in unit of the
      inverse of the pixel size multiplied by $2\pi$, corresponding
      roughly to $k \simeq 1/L({\rm Mpc})$. {\it Second row of
      panels:} the probability distribution function as a function of
      $\gamma_{\HI}=\ln(1+\delta_{\HI}$ 
      (as it is indicated at the bottom), after smoothing
      $\gamma_{\HI}$  with a Gaussian window of size $\ls= \sqrt{2}
      \dlos$.  {\it Solid thick and dashed  thin} lines correspond to
      the recovered and the original fields, respectively. The 
      light shaded region in each panel represents the scatter derived from
      the five GRFs,  while the big dots correspond to gaussian
      profiles with same mean and same variance as the smoothed
      recovered fields.  {\it Third row of panels:} similarly as for
      the second row, but for  the Euler characteristic.  
      {\it Fourth row of panels:} similarly as for
      the third row, but  for the individual critical point counts. In
      that case, the {\it thick and thin lines} correspond to the
      recovered and original fields, respectively. The {\it solid,
      dashed, dotted and dot-dashed lines}  correspond respectively
      to minima, pancake saddle points, filament saddle points and
      maxima.}
    \label{fig:power_spectrum_LOG}
   \end{figure*}
  \begin{figure*}
    \centering
    \subfigure[]{\includegraphics[width=6.cm]{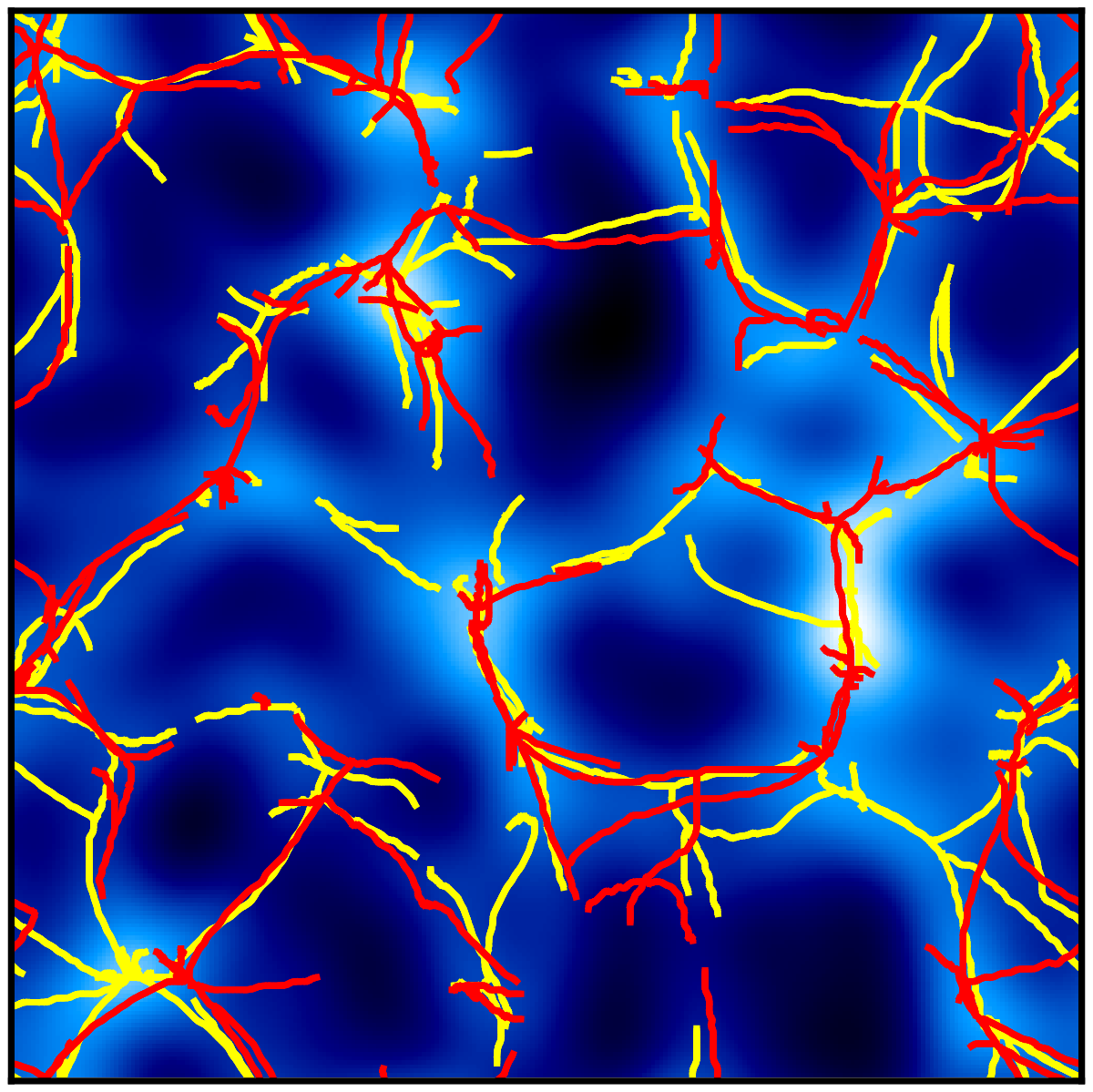}
      \label{fig:skl_pic}}
    \hspace{0.6cm}
    \subfigure[]{\includegraphics[width=6.cm]{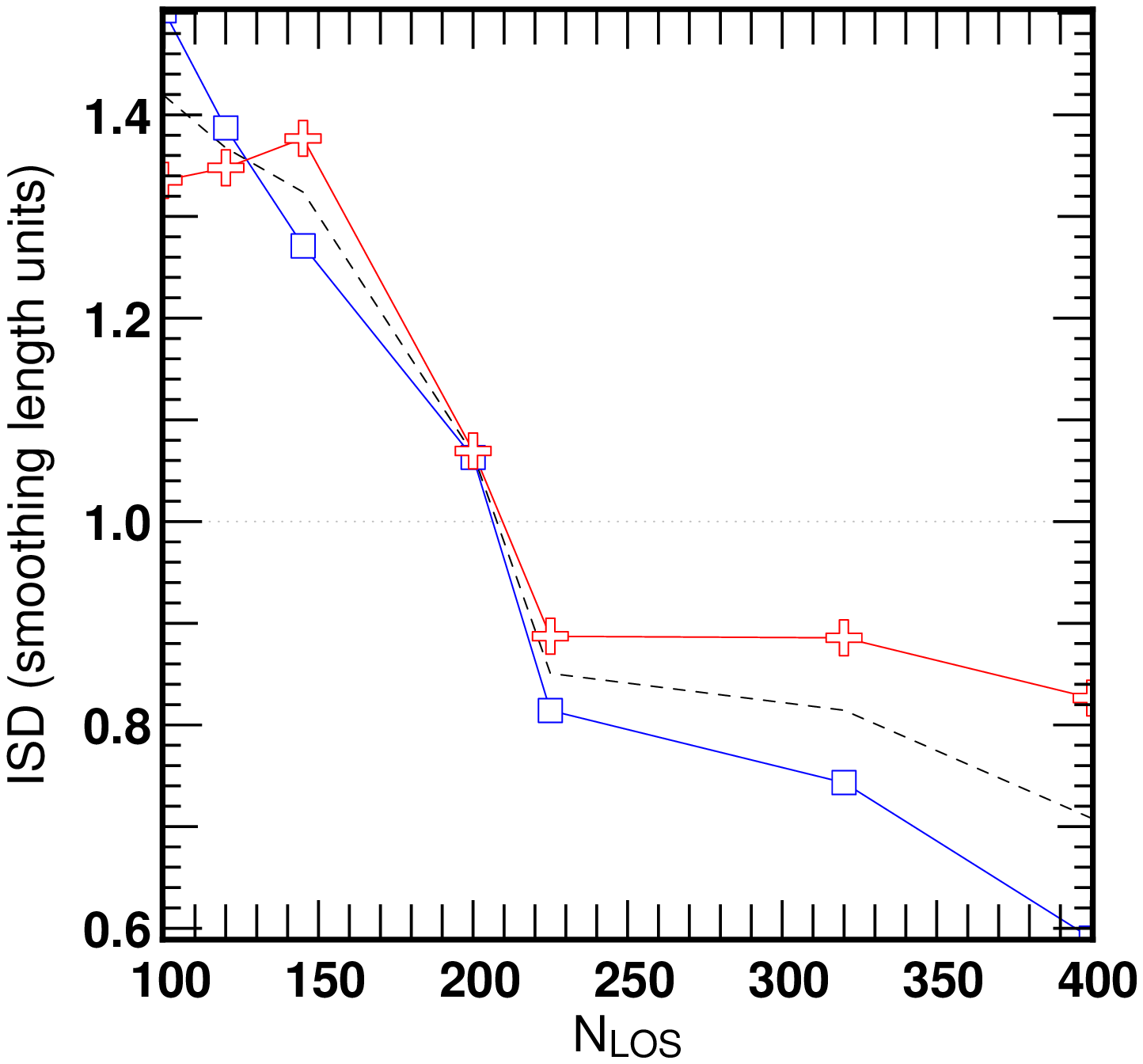}
      \label{fig:skl_dist}}
    \caption{ {\it Left panel:}  Comparison between the skeletons of
      the original field {\it (light lines)} and its recovered counterpart
      {\it (darker lines)}  (the skeletons  represented here are the true
      ones, and not their local  approximations, as defined in  the
      main text).  The original field was recovered by inverting
      $\nlos=320$ \loss,  corresponding to a separation $\dlos = 2.24$
      Mpc. Both skeletons are computed on  fields smoothed over a
      scale $L_{\rm s} = 3.16$ Mpc, in logarithmic space. For clarity,
      only a $4$ Mpc slice is shown, the background contour
      representing the original smoothed density field (lighter colors
      corresponding to higher densities).  {\it Right panel:}
      Evolution of the inter-skeleton distance (ISD) between the
      original and  reconstructed fields as a function of the number
      of line of sight $N_{\rm LOS}$.  The ISD is computed after
      smoothing over a scale $L_{\rm s} = 3.65$ Mpc which is roughly
      equivalent to the lowest resolution reconstruction sample. The
      upper {\it (crosses)} and lower {\it (squares)} curves represent the
      measured median distance from the  reconstructed field skeleton
      to the original one and vice versa respectively, while the
      dotted curve represent their average value.}
      \label{fig:skel}
  \end{figure*}

First note that the inversion is not directly performed on the
density, but on its logarithm, i.e.  the normalized density field is
written as $\rho_{\HI}=1+\delta_{\HI}\equiv \exp(\gamma_{\HI})$ and
the field $\gamma_{\HI}$ is interpolated by using the method  described
above.  This has two advantages:  (i) it ensures the
positivity of the reconstructed density; (ii) since the density turns
out to be roughly log-normal\footnote{Note that if a field such as
$\rho_{\rm gas}$ is lognormal, the (inverse of the) transformation
(\ref{eq:powerlaw}) leaves the  new field, e.g. $\rho_{\HI}$,
lognormal as well.} (see for example  Bi \& Davidsen, 1997;
Choudhury, Padmanabhan \& Srianand, 2001; Viel \etal, 2002a; Zaroubi
\etal; 2006, Desjacques \etal, 2007; and see  Coles \& Jones, 1991,
for the statistical properties of the log-normal distribution),
performing the reconstruction on the logarithm of the field is
expected to reproduce more realistic results as shown on
Fig.~\ref{fig:PDFs_densfields}.

  However, as a result of the reconstruction, the recovered field, 
  $\gamma_{\HI,\rm rec}$, will be smooth over anisotropic volumes of
  size $\sim L_x \times L_T \times L_T$,  which means that at best,
  one can identify structures at this level of  smoothness on a
  logarithmic space. Although  theoretical predictions (namely
  gravitational  clustering,  primordial non Gaussianities, etc. )
  do exist for the density field itself, that is for $\rho_{\HI}=\exp(
  \gamma_{\HI})$ and its smoothed  counterparts, they can not be
  applied directly in our case because smoothing and
  taking the exponential are operations which do  not commute, except
  in the weakly nonlinear regime,  $\delta_{\HI} \ll 1$. In
  particular, recovering the results for $\delta_{\HI}$
  on a  linear space, by taking
  the exponential of  $\gamma_\HI$ and subsequently smoothing
  it,  an effective bias, essentially due to rare peaks in the
  $\gamma_{\HI,\rm rec}$ field,  is introduced. The effect of such a
  nonlinear bias is difficult to control and can in  some cases be
  important as shown below and studied in more details in
  Appendix~\ref{appendix}.

   \begin{figure}
     \centerline{\psfig{width={7.cm},figure=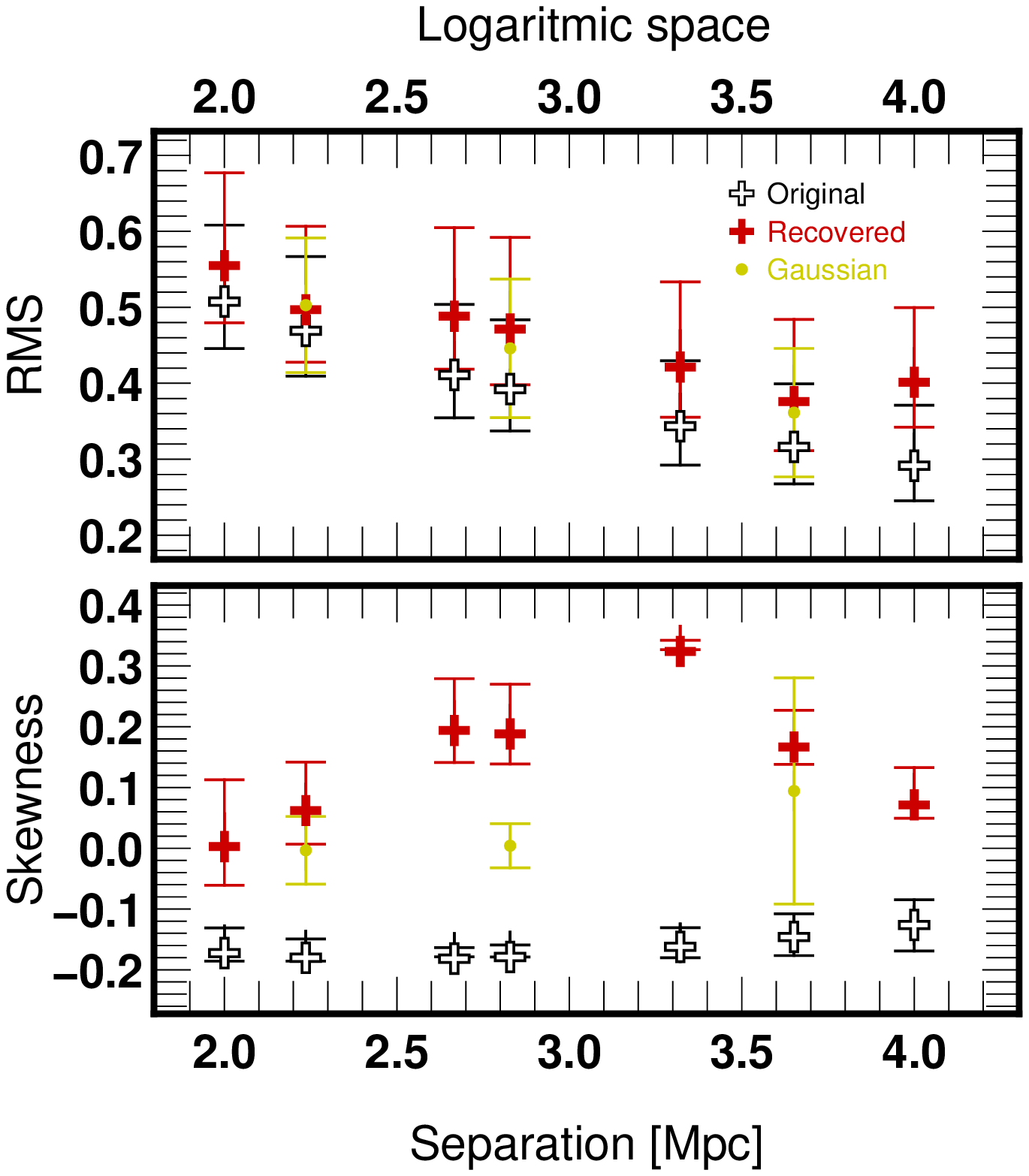}}
     \caption{Variance {\it (top panel)} and skewness {\it (bottom
      panel)} of $\gamma_{\HI}=\ln(1+\delta_{\HI})$ for the original
      (open crosses), the recovered (filled crosses) fields and the
      Gaussian prediction (light dots), as functions of the LOS
      separation, $\dlos$.  The symbols correspond to measurements
      performed on the $\gamma$ fields smoothed with a Gaussian window
      of size  $\ls=\sqrt{2}\dlos$. For a proxy of the errorbars, we
      used the measurements at smoothing scales $\ls=\dlos$ and
      $\ls=\sqrt{3}\dlos$, except for the Gaussian fields, where the
      dispersion among the 5 realizations is a better choice.  For the
      Gaussian case, the skewness should be exactly zero.  The
      measurements are consistent with that value, despite the large
      dispersion at the largest scales, due to finite volume effects.
      }
     \label{fig:skewness_rms_recorig_LOG}
   \end{figure}
  \begin{figure}
    \centerline{\psfig{width={7.cm},figure=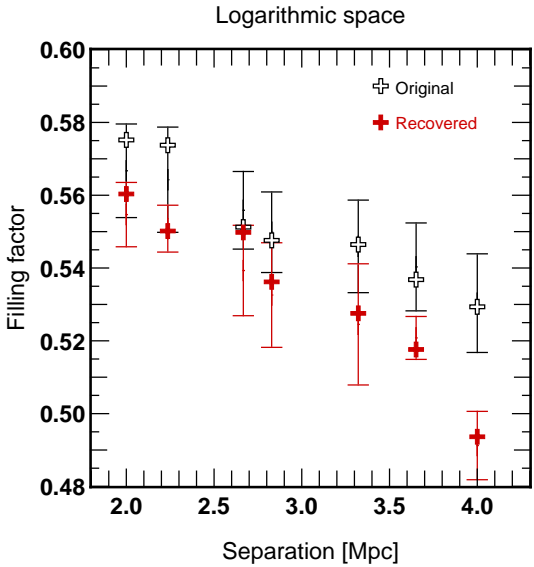}}
    \caption{Same as in Fig.~\ref{fig:skewness_rms_recorig_LOG} but for the 
      filling factor of underdense regions at the minimum of
      the Euler characteristic. The Gaussian limit, not shown here,
      would give a filling factor exactly equal to 0.5.
      }
    \label{fig:FFvoids_rec_orig}
  \end{figure}
  \begin{figure}
    \centerline{\psfig{width={7cm},figure=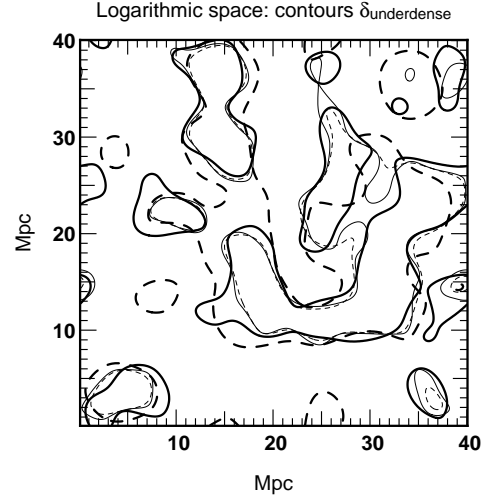}}
   \caption{Contours of underdense regions estimated from the  minimum
      of the Euler characteristic in logarithmic space.  The  thick
      curves represent the contours for two recovered fields
      $\gamma_{\HI,{\rm rec}}$ obtained by the inversion of
      $\nlos=320$ and $\nlos=200$ \loss\ (solid and dashed lines
      respectively).  Prior to contour determination, the recovered
      field was smoothed with a Gaussian window of size $\ls= \sqrt{2}
      \dlos$.  These contours should be compared with those of the
      original field, $\gamma_{\HI}$, represented with a thin line,
      smoothed at the same scales (solid and dashed lines
      respectively).  This figure is complemented  with the two upper
      panels of the lower left group shown in
      Fig.~\ref{fig:slice_recorig}, where the same slices for the
      original and the recovered fields are displayed.  }
    \label{fig:contour_voids}
  \end{figure}
 \begin{figure}
    \centerline{\psfig{width={6.7cm},figure=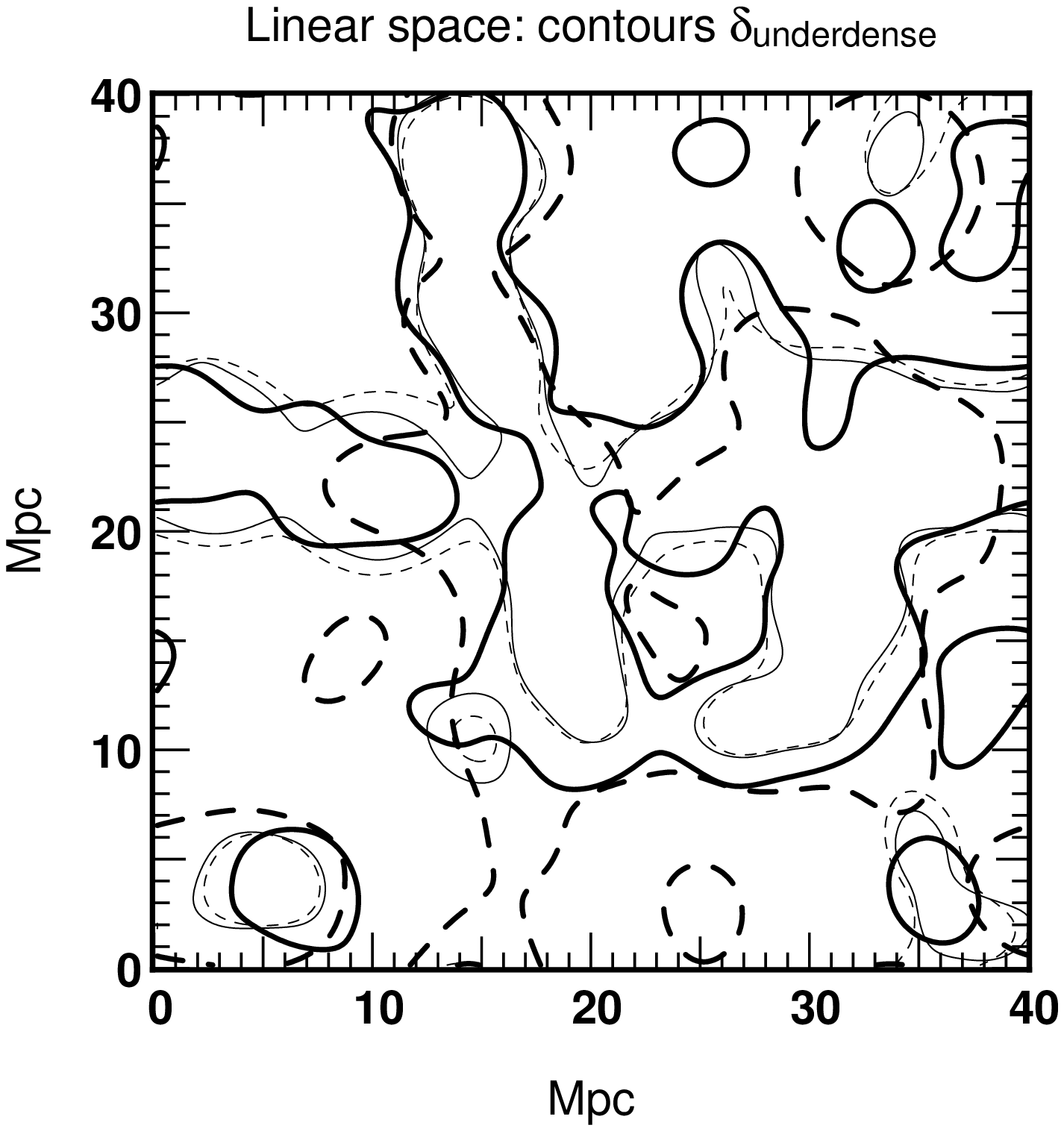}}
    \caption{Same as Fig.~\ref{fig:contour_voids}, but smoothing is
    performed in linear space. This figure is complemented  with the
      two upper panels of the lower right group  
      in Fig.~\ref{fig:slice_recorig}, where the same slices for 
      the original and the recovered fields are displayed.
    }
    \label{fig:contour_voids_exp}
  \end{figure}

  % ======================================================================== 
\subsection{Testing the reconstruction: statistical and topological  analysis}
  \label{sec:euler_recov_fields}
  % ========================================================================

  We now test the quality of the reconstruction using the same
  statistical tools as in Section \ref{sec:euler},  namely the
  PDF of the field and the Euler
  characteristic.  Other statistics are considered, such as the
  variance and the  skewness of the PDF, the power-spectrum of the
  density field and the filling factor of regions less dense than the
  minimum of the Euler characteristic.  In addition, to have a
  quantitative estimate of the accuracy in the locus of the
  filamentary structures, we use the skeleton as introduced  by
  Novikov, Colombi \& Dor\'e (2006) and by Sousbie et al. (2007) and
  define an inter-skeleton distance (ISD).

  Following the discussion in Section~\ref{sec:biasrec}, the
  reconstruction is mainly tested on the field $\gamma_\HI$
  and its smoothed counterparts.  In Appendix~\ref{appendix} 
  we provide additional results on the field $\delta_{\HI}$.

  Since there are two scales in the inversion (see \Sec{paraminv}), the
  recovered optical depth is an anisotropic smooth field with fewer
  structures in the direction transverse to the  LOSs than in the
  direction parallel to them. Optimal comparison between
  reconstructed and real optical depth would require an approach based
  on anisotropic smoothing, a level of complexity well beyond the
  scope of this paper.  Instead, to compare the reconstructions to the
  exact solution, an isotropic smoothing via a Gaussian window is used
  (see Eq.~\ref{eq:smoothingwindow}).
  The width of the smoothing window is $L_{\rm s} \ga \langle
  d_{\rm LOS}\rangle= L_T$. The choice of the optimal smoothing scale
  is constrained by the inter-skeleton distance.
  
  One of the uncertainties in the reconstruction involves the
  determination of the mean value of the field  $\mu_{\rm true} \equiv
  \langle \gamma_{\HI} \rangle \equiv  \langle \ln \rho_{\HI}
  \rangle$, which can in principle be estimated only along the LOSs.
  To improve the quality of the reconstruction,  its average is fixed
  to  $\mu_{\rm true}$:\footnote{Note that the inversion formula,
  equation~(\ref{eq:inversioncombo}) can  be amended  to impose
  directly this constraint, following equation~(11) of PVRCP, by
  including this  information in  ${\bf M}_0$.}\footnote{When the
  analyses are performed in linear space, the normalization is
  different, as discussed in Appendix~\ref{appendix}.}
  \begin{equation}
    \gamma_{\HI,\rm rec}={\tilde \gamma}_{\HI, \rm rec}- \langle
    {\tilde \gamma}_{\HI, \rm rec} \rangle + \mu_{\rm true}.
    \label{eq:normrho}
  \end{equation}
  In practice, the knowledge of $\mu_{\rm true}$ is expected to be
  accurate, even though its actual measured value, $\mu_{\rm LOS}$, is
  determined  along the LOSs. For instance in the worse  case
  considered in this work,  $N_{\rm LOS}=100$,  $\langle (\mu_{\rm
  LOS}- \mu_{\rm true})^2 \rangle^{1/2}/\vert \mu_{\rm true}\vert
  \simeq 1.91\%$, where the mean value of the difference between the
  measured and the real  $\mu$ has been calculated by averaging over 100
  different realizations of 100 LOSs.

 % ==========================================================================
 \subsubsection{Visual inspection}\label{sec:visu}
 %  ==========================================================================
  A first qualitative comparison between the  original and the
  recovered fields in logarithmic space can be made by examining
  \Fig{slice_recorig}. The top panels illustrate  the  anisotropic
  nature of the reconstruction.   Smoothing at a scale $L_{\rm
  s} \ga \langle d_{\rm LOS} \rangle$ (for example, in the  case of
  Fig.~\ref{fig:slice_recorig}, $\ls=\sqrt{2}\dlos$), greatly improves
  the agreement   between the reconstruction and the exact solution 
  and the two field become almost indistinguishable
  (bottom left panels). When one examines in detail where the
  reconstruction fails, one notices that these structures  correspond
  to overdense regions. The fine nature of the web formed by overdense
  regions (filaments, clusters) makes the reconstruction more
  difficult for these regions than for the underdense ones because of
  the sparse sampling of the transverse structures.

  When going to linear space, i.e. taking the exponential of the
  fields and subsequently smoothing them,  the effect caused by the
  amplification of rare events discussed in \S~\ref{sec:biasrec}
  becomes obvious, as illustrated by the bottom right panels of
  \Fig{slice_recorig},  that represent the counterpart of the bottom
  left panels in linear space. In logarithmic space the highest
  density peaks are highly depleted, here they are  visible and
  spread over a beam the  typical size of which is that  of the smoothing
  window.

%==============================================================================
\subsubsection{Power-spectrum: the scales correctly reconstructed}  
%==============================================================================
The good agreement between the original and the recovered fields in
logarithmic space is confirmed by the first row of panels in
Fig.~\ref{fig:power_spectrum_LOG},   which shows the power-spectrum,
$P(k)=\langle |\gamma_{\bf k}|^2 \rangle$,  of the raw fields,
$\gamma_{\HI}$ and $\gamma_{\HI,\rm rec}$, for three reconstructions
($N_{\rm LOS}=320,\ 200,\ 120$, corresponding respectively to $\langle
d_{\rm LOS} \rangle=2.24$, $2.83$ and $3.65$ Mpc). We also show  five
realizations of a  Gaussian random field (GRF) with the same $P(k)$ as
$\gamma_{\HI,\rm rec}$, in order to estimate finite volume effects.  As
expected, the filtering nature of the reconstruction introduces
an apodization effect on  $P(k)$ visible on
Fig.~\ref{fig:power_spectrum_LOG}: a bending of $P(k)$ is expected to
happen roughly for $k \simeq k_{\rm bend} \equiv 2\pi/L_{\rm T}$, i.e.,
$k_{\rm bend}=0.44$, $0.35$, $0.27$ from the upper left to the upper
right panel, respectively, in the units chosen in
Fig.~\ref{fig:power_spectrum_LOG}. It is not straightforward to check
accurately this property by visual inspection.  Indeed, when $N_{\rm
LOS}$ decreases, the small $k$ part of the reconstructed
power-spectrum becomes less well correlated with the true $P(k)$,
giving the illusion, for example, that  overall  the $N_{\rm LOS}=120$
reconstruction does better than the $N_{\rm LOS}=320$ one.  Still for
$k \la k_{\rm bend}$, i.e., 
$L_{\rm s} \ga L_{\rm T}=\langle d_{\rm LOS} \rangle$, there is 
a good agreement between the reconstruction and the exact
solution. However, the measurement of the power-spectrum itself is not
accurate enough neither does it contain enough information to guarantee
that filaments are well reconstructed in detail, as we examine now.

% ===========================================================================
\subsubsection{The skeleton: optimal smoothing scale}
\label{sec:myske}
% ============================================================================
The visual inspection of \Fig{slice_recorig} seems to show that the
filamentary pattern of the overall three-dimensional distribution is
well recovered by the reconstruction in logarithmic space. One can
check that assertion more quantitatively on the skeleton (Novikov
\etal, 2006; Sousbie \etal, 2007). This will allow us  to define
an optimal smoothing scale which will be 
used in the subsequent analyses. More
detailed analyses relying on the skeleton are postponed to another
paper.

The actual definition of the skeleton is in fact  deeply related to
the Euler characteristic since it relies on first principles of Morse
theory: basically, the skeleton is composed of the set of field lines
(the curves defined by the gradient of the density field) starting
from the filament saddle points ($I=2$ in the formalism described in
\S~\ref{sec:euldef}) and converging to local maxima.\footnote{The
actual conditions for this definition to be valid are discussed in
Novikov \etal\ (2006).}   Although apparently simple, solving this
equation is quite difficult, this is why a local approximation, based
on a Taylor expansion around the critical points contained in the
skeleton, was introduced in  Novikov \etal\ (2006) and extended in 3D
by Sousbie \etal\ (2007):  the {\em local} skeleton. In this paper we
use the implementation of Sousbie \etal\ (2007).

Note that, as opposed to a global topological estimator such as the
Euler characteristic, the skeleton provides a {\it local} test of the
accuracy of reconstruction (i.e., one can check whether a precise
filament at a given location  is recovered or not). Figure
\ref{fig:skl_pic} presents the skeleton  (yellow lines) of a $4$ Mpc
slice extracted  from the original \HI\ density field, as well as the
skeleton of its reconstructed counterpart (red lines),  using $320$
LOSs. Both fields are smoothed  in logarithmic space using a gaussian
window of scale $L_{\rm s} = 3.16$ Mpc. This figure confirms that, on
large scales, the general shape of the filamentary structures is  well
preserved,  demonstrating the ability of the reconstruction to recover
the cosmic web. Nonetheless, as expected, some  discrepancies appear
on small scales.

The inter-skeleton distance (ISD) is an estimator which allows  to
make a quantitative comparison.  A skeleton corresponds to a number of
small segments  linked together to form the filaments.  In order to
estimate the average distance between two skeletons, A and B, for each
segment of A, the  distance to the closest segment in B is computed
leading to the PDF of the distribution of the spatial separation
 from A to B. The
distance from A to B is defined as the median of this PDF.  Since this
definition of distance is not symmetrical (in the sense that ISD(A,B)
and ISD(B,A) will, in general, be different), the mean distance
between A and B is defined as the average of  these quantities. Figure
\ref{fig:skl_dist} presents the measurement of the ISD between the
skeleton of the original field and its reconstruction, as a function
of the number of LOSs used to perform the inversion (in units of the
smoothing scale). In all  cases, both fields were smoothed over a
scale  $L_{\rm s} = 3.65$ Mpc.  What is important to notice here is
the sharp transition at  $N_{\rm LOS}=225$, corresponding to $L_{\rm
s} = L_{\rm crit}$ with
\begin{equation} 
L_{\rm crit} \simeq 1.35 \langle d_{\rm LOS} \rangle.\label{eq:lcrit}
\end{equation} 
For a smoothing scale $L_{\rm s} \la L_{\rm crit}$, the match between
reconstruction and exact solution worsens suddenly, while no
significant improvement is really seen when $L_{\rm s} \ga L_{\rm
crit}$: $L_{\rm crit}$ represents some ``optimal'' smoothing scale,
which is the smallest possible scale at which the reconstruction
performs well, in terms of filamentary pattern recovery.  Note that
only the measurements for a particular value of $L_{\rm s}$ are shown,
but Eq.~(\ref{eq:lcrit}) should not change significantly for the
scaling range considered in this work.

Although all the subsequent analyses involving smoothing were
performed at various scales, namely $L_{\rm s}^2=\langle d_{\rm LOS}
\rangle^2$, $2 \langle d_{\rm LOS} \rangle^2$, $3  \langle d_{\rm LOS}
\rangle^2$, increasing likewise the number of LOSs contributing per
smoothing volume, we shall, in light of the above findings, mainly
concentrate on the results corresponding to $L_{\rm s}^2=2 \langle
d_{\rm LOS} \rangle^2 \simeq L_{\rm crit}^2$.

%==============================================================================
\subsubsection{Statistical analysis}
%==============================================================================
The second row of Fig.~\ref{fig:power_spectrum_LOG} shows
the PDFs of  the smoothed counterparts of $\gamma_{\HI}$ (dashes) and
$\gamma_{\HI,\rm rec}$ (solid), with a window of size $L_{\rm
s}=\sqrt{2} \langle d_{\rm LOS} \rangle$, as argued just above.  These
measurements are supplemented with
Fig.~\ref{fig:skewness_rms_recorig_LOG},  which shows the variance and
the skewness of the PDFs of various fields as functions of separation
between the LOSs. Again, the  agreement between the solid and dashed
curves in second row of panels of Fig.~\ref{fig:power_spectrum_LOG} is
quite good and the results do not depend significantly on the value of
$N_{\rm LOS}$. 

 From a quantitative point of view, 
the difference between  the recovered and the original curves can
be calculated using the following estimator,
\begin{equation}\label{eq:estim-error}
  {\rm err} = \frac{\sum_i |y_i^{\rm orig}-y_i^{\rm rec}|\Delta x_i}
  {\sum_i |y_i^{\rm orig}|\Delta x_i}\,,
\label{eq:diffint}
\end{equation}
where $y_i^{\rm orig}=y^{\rm orig}(x_i)$ and $y_i^{\rm rec}=y^{\rm
rec}(x_i)$ are the values of the curves relative to  the original and
the recovered fields respectively and the  curves have been sampled at
points  $x_i$.  This corresponds to  the area between the curves,
normalized by the area enclosed by the original ones.  
For the three reconstructions shown, 
the errors are of the order of ${\rm err}_{\rm PDF}=10-20\%$.

 These quantitative estimates show that there
are still some noticeable differences between the
reconstruction and the true field: the shape of PDF of the
reconstructed field, $\gamma_{\HI,\rm rec}$, tends to be Gaussian,
within the error range provided by the five Gaussian fields.  This
``Gaussianisation'' is expected from both the central limit theorem
and the shape of the correlation matrix given by
Eq.~(\ref{eq:correl_matrix}).  Note that this statement is not totally
consistent with the measurement of the skewness (lower panel of
Fig.~\ref{fig:skewness_rms_recorig_LOG}), especially at intermediate
separations between the LOSs. However, this skewness is  quite
sensitive to the upper tail of the PDF corresponding to rare events in
overdense regions: one expects, in that regime, deviations from
Gaussianity in the reconstruction because the central limit is  not
yet reached.

  The true field, $\gamma_{\HI}$, deviates slightly  from a Gaussian,
  as already shown in Fig.~\ref{fig:PDFs_densfields}. In particular,
  in the right part of the bell shape of the PDF on
  Fig.~\ref{fig:power_spectrum_LOG}, there is a slight disagreement
  between the dashed and the continuous curves, which corresponds  to
  the weak negative skewness measured in lower panel of
  Fig.~\ref{fig:skewness_rms_recorig_LOG}. This disagreement would be
  even more visible if a logarithmic representation were used on the
  $y$ axis to display the PDF: the high density tail of the \HI \
  field is far from lognormal. The main contribution to such a tail
  comes from collapsed objects in clusters and in filaments. As argued
  in \S~\ref{sec:visu}, these objects are sparsely sampled by the
  LOSs, which worsens the quality of the reconstruction in overdense
  regions.

%======================================================================== 
\subsubsection{Global topology}  
%========================================================================
The nearly Gaussian nature of the reconstructed $\gamma$ field can be
  also confirmed  by examining the third row of panels in
  Fig.~\ref{fig:power_spectrum_LOG}, which is similar to the second
  row, but displays the Euler characteristic as a function of the
  density threshold.  Deviations from Gaussianity of the true field,
  $\gamma_{\HI}$, are more clearly visible than for the PDF. In
  particular, on all the panels, the corresponding dashed curve always
  presents an asymmetry  between its two maxima, contrary to what is
  observed in the   Gaussian limit. 
  The reconstruction, $\gamma_{\HI,\rm rec}$, being
  more symmetrical,  is clearly closer to the Gaussian limit than
  the true field.  However, as noticed earlier for the skewness of the
  PDF, one cannot really claim that the reconstruction  is fully
  Gaussian: deviations outside the range allowed by our five Gaussian
  realizations are noticeable, particularly in the right panel and in
  general in the overdense right tail ($\gamma \ga -1.5$) of
  $\chi^+$. Still the overall topology of the reconstructed field,
  although closer to the Gaussian limit,  reproduces  rather well the
  topology of the true field, especially in the range $-2.0 \la \gamma
  \la -1.5$, which confirms the findings of \S~\ref{sec:myske} on the
  skeleton. This density regime is indeed dominated by filaments
  saddle points and local maxima, as shown by the last row of panels
  of Fig.~\ref{fig:power_spectrum_LOG}, which displays  the  different critical
  point counts as  functions of the density  threshold for
  $\gamma_{\HI}$ and $\gamma_{\HI,{\rm rec}}$ .  Note the
  increasing contribution of the noise when $N_{\rm LOS}$ decreases,
  which makes the agreement between reconstruction and exact solution
  worse, particularly for large densities, as expected.
   From a more quantitative point of view, one can, similarly
  as for the PDF, compute the integrated errors on the critical point
  counts (Eq.~\ref{eq:diffint}). For the 3
  reconstructions we consider here, 
  these errors are of the same order as for the PDF 
  (i.e. less than 20\%).

  As an additional test, the minimum of the Euler characteristic,
  $\gamma_{\rm min} \sim -2$, can be used to define a topological
  boundary between ``underdense'' and ``overdense'' regions. Indeed in
  the Gaussian limit, this minimum lies exactly at $\gamma_{\rm
  min}=\langle \gamma \rangle$. Defining the filling factor of
  underdense regions, $FF_{\rm  underdense}$, as the fraction of space
  occupied by points verifying $\gamma \leq \gamma_{\rm min}$, one
  expects $FF_{\rm underdense}$ to be always close to $0.5$: at least
  this is true for any monotonic local transform of a Gaussian field
  (with no additional smoothing).  Even though the reconstruction and
  the true field do not have exactly the same behavior for $\chi^+$,
  they seem to have very close values of $\gamma_{\rm min}$, which
  should correspond to a good agreement between the measured values of
  $FF_{\rm  underdense}$: this is indeed the case as shown in
  Fig.~\ref{fig:FFvoids_rec_orig}.  Although the measured values of
  $FF_{\rm  underdense}$ are consistent with those of the original
  ones,  the connectivity of the underdense (or equivalently,
  overdense) regions defined in this way is good but not perfect, as
  illustrated by Fig.~\ref{fig:contour_voids}. In this  range of densities,
  connectivity of the excursion is controlled equally by filament and
  pancake saddle points and their respective counts tend to be
  slightly underestimated  and overestimated, respectively, as
  illustrated by bottom panels of
  Fig.~\ref{fig:power_spectrum_LOG}. This is however not enough to
  explain the discrepancies in Fig.~\ref{fig:FFvoids_rec_orig}, and
  shows the limits of global topological estimators.

  At the qualitative level, note finally that the situation becomes
  worse when one  attempts to recover the boundary contour between
  overdense and underdense regions in linear space, because of the bias
  mentioned in \S~\ref{sec:biasrec}. This is shown in
  \Fig{contour_voids_exp}, which represents the same slice  as in
  \Fig{contour_voids} but in this case the contours were calculated
  from the minimum of the Euler characteristic after smoothing the
  exponential of the fields.  Here, the position of the structures in
  the recovered contours is significantly different from that of the
  original fields, not to mention connectivity.  Appendix
  \ref{appendix}, which discusses a figure similar to
  Fig.~\ref{fig:power_spectrum_LOG} but in linear space, fully
  confirms these results.

%======================================================================
\section{Discussion and conclusion}\label{sec:conclusion}
%======================================================================

   In this paper we have studied the topology of large scale
  structures as traced by the intergalactic medium (IGM) in a
  hydrodynamical cosmological simulation. The main goal was to test a
   reconstruction method (PVRCP) of the three-dimensional large scale matter
  distribution from multiple lines of sights (LOSs) towards
  quasars. For this purpose,  we  relied on a number of global
  statistical tools, the probability distribution function of density
  field (PDF), the Euler characteristic ($\chi^+$) as an alternate
  critical point counts and related quantities such as the variance and
  the skewness of the PDF, the individual critical points counts and
  the filling factor of the underdense regions at the minimum of the
  Euler characteristic.  We also used the skeleton as local probe of
  the geometry  and the topology of the field. The main
  results of our investigations can be summarized as follows
  \begin{itemize}
  \item In the first part of this paper we addressed the problem of relating
  the topology of the dark matter density field to the topology of the
  distribution traced by the total amount of gas and the neutral
  gas (\HI).  When one considers the \HI\ density distribution at
  scales larger than the Jeans length of the gas and takes into
  account the IGM equation of state relating the neutral and total
  amount of gas, then the properties of this  nearly
  lognormal distribution are exactly
  the same as found for the dark matter/total gas in underdense
  regions (i.e.  for density contrasts $\delta\la 0$).  For larger
  density contrasts, some deviations appear, due to shocks (where \HI\
  is depleted) and to the presence in filaments  and clusters of
  highly condensed objects (where \HI\ is very concentrated).  Taking
  these results into account, with the additional assumption that
  instrumental noise,  in particular effects of saturation, can be
  neglected, we have shown   that studying the topological properties
  of large scale matter density distribution is equivalent to studying
  directly those of the optical depth or in what follows, those of
  neutral gas, \HI.

  \item In the second part of this work we tested the Wiener interpolation
  proposed by PVRCP to recover the three-dimensional distribution of
  \HI\ from a set of multiple LOSs, along which the (one-dimensional)
  distribution of \HI\ is assumed to be  known exactly.  This
  interpolation depends on three parameters, the typical size, $L_x$
  of structures along the LOSs,  the typical mean LOSs separation,
  $L_T=\dlos$, and the expected variance of the fluctuations of the
  field   which can be in principle indirectly inferred from the LOSs
  themselves.

  Our investigation shows that the reconstruction method can be used
 to predict  quite accurately the  patterns in the large scale matter
 distribution at scales  of the order of $\sim 1.4\dlos$ or larger
 when one attempts to recover the {\it logarithm} of the  density
 field.  In particular it allows us to recover the position of
 filaments in the large scale distribution:
  we compared the skeleton of the initial and recovered field and measured
 the distance between these skeletons and found that  for smoothing
 scales larger than $\sim 1.4\dlos$, the inter-skeleton distance
 remains smaller than $\dlos$.  Furthermore,  the  global
 shape of the PDF, of the fraction of critical points and of  the
 Euler characteristic are well reproduced,  the integral errors
 on these quantities varying in the range 10-20\%. Discrepancies
 between the reconstruction and the exact solution are mainly found
 in overdense regions, where deviations from a lognormal behavior
 are the most significant.
 \end{itemize}

 The good recovery of the statistical
properties of the density field in  logarithmic space,
is strongly related to  the Gaussian 
prior on which the inversion method is based. 
Recall that, since the distribution of the gas density is very close to
log-normal, the distribution of 
its logarithm is well approximated by a Gaussian 
function. As demonstrated in PVRCP, the 
 Wiener interpolation is just a special case of the maximum 
likelihood method. It gives, under the hypothesis that the 
statistical distributions of the data and of the parameters are Gaussian,
the optimal reconstruction for a linear model. However, this relies on a proper
knowledge of the covariances matrices.
Here we assume a simple functional shape for these matrices,
given by \Eq{correl_matrix}. A better
treatment would need an accurate knowledge of the underlying
power-spectrum of the logarithm of the density. The interpolation
could for instance be improved by using a stronger prior relying
on the extension of e.g. the 
nonlinear ansatz of Hamilton et al. (1991) to logarithmic space.

We noticed that some deviations are present in  the
original field, compared to the log-normal limit at the scales we have
probed here. This information could be added to the model.
This could be achieved by applying an Edgeworth expansion to the 
logarithm of the field \cite{jus95,skewedLN}, hence
by taking into account slight deviations of the likelyhood
function from a Gaussian distribution to  correct our
Wiener interpolator (Amendola, 1998). 

Even though the best variable for the reconstruction is  the
logarithm of the density,  theoretical predictions are usually
performed on the density itself. Therefore it is in practice
difficult to compare the properties of the 
reconstructed density distribution
to those predicted by e.g. nonlinear perturbation theory 
(e.g. Bernardeau et al. 2002) or other models. The
problem is that linear space  gives more  emphasis to rare
events in overdense regions.  In Appendix~A we analyse the
corresponding bias on the reconstruction, 
and find that it  is critical for the higher
density peaks.  As a result, the tomography is in practice much 
less robust when expressed directly in linear space. However,
this is mainly related to the fact that our analyses are performed
at scales smaller or of the order
of 4 Mpc, where non-linear effects in the dynamics are still present.

Due to the size of our simulation ($L_{\rm box} = 40$ Mpc),
in this work we have analysed the properties of connectivity at
relatively small scales ($\ls = 4$ Mpc), where the distribution of
matter is close to log-normal\footnote{Note that,
because of the small size of our simulation, we could not
really examine the effects of cosmic variance, except
with our Gaussian realizations.}.
However,  one could in principle extend the analyses to larger scales,
to probe the linear or quasi-linear regime, where the density
distribution is actually close to Gaussian. In that case,
the reconstruction should be performed on the density
itself rather than on its logarithm while  the above mentioned  problems
would become irrelevant. In particular, the implementation of the
improvements on the Wiener interpolator could for instance be used 
to test directly  if  non trivial  deviations from Gaussianity are 
present or not in the data. If present, they could  be ascribed 
to primordial non-gaussian features that are  produced during 
the inflationary phase or  as a  result of topological defects.

The inversion method is based on the 
hypothesis that a sufficiently strong correlation exists at the scale 
under consideration. Indeed, various sources
of noise can hide such a correlation completely 
(errors due to the finite cosmological volume probed by a finite 
number of LOSs, noise in the measurement of the spectra),
making the reconstruction irrelevant. 
To test the strength of the 
correlation a large number of quasar pairs spanning 
the range of separations we want to probe must be observed. It has 
been recently shown \cite{coppolani06} that at $z\approx 2$ for a 
separation of $\sim 5$ arcmin (corresponding to $\approx 7.6$ Mpc for 
a flat universe with $\Omega_{\rm m}=0.3$, $\Omega_\Lambda=0.7$ and
$H_0=70\,{\rm km}\,{\rm s}^{-1}\,{\rm Mpc}^{-1}$), some correlation is 
observed, suggesting that the inversion method could be applied
at such scales. It is thus very important to measure more accurately 
the transverse correlation 
function from quasar pairs. Indeed, once this is done, 
we can include this information as a self-consistent prior 
in the reconstruction procedure.

Using realistic data about the luminosity function of 
quasars \cite{lumfct_QSO},
it is found that for magnitude limits of $g\lsim(23,24,25)$ the number
of quasars observed per square degree at $z\gsim 2$ is 
$n_{\rm QSOs}=(41,77,136)$ respectively. For the set of cosmological 
parameters assumed here, the corresponding mean angular separations  
are $\dlos=(9.37,6.84,5.15)$ arcmin.
Moreover, for $g\gsim 23$ the number density of Lyman-break galaxies
(LBGs) starts to become significant and we can think of using these
objects as background sources in combination with QSOs.
In particular, it is found that for $g\lsim(23,24,25)$ the number
of LBGs par square degree is $n_{\rm LBGs}=(0.3,116,2325)$  
respectively \cite{lumfct_LBG}, so that, 
even at  $g\lsim 24$, the number of available 
sources is largely increased. In Table \ref{tab:qso_lbg} we display 
 the mean separation one can expect as a function of 
the magnitude limit.

\begin{table}
  \begin{center}
    \begin{tabular}{|c|c|c|}
      \hline
      Magnitude limit  & Separation (QSOs) & 
      Separation  (QSOs et LBGs)\\
      $g$ & (arcmin) & (arcmin)\\
      \hline  \hline
      23  & 9.4  & 9.3  \\
       24  & 6.8  & 4.3  \\
       25  & 5.2  & 1.2  \\
      \hline
    \end{tabular}
  \end{center}
  \caption{Mean angular separation between the background sources 
    as a function   of the magnitude limit (left column).
  }
  \label{tab:qso_lbg}
\end{table}

One can see that if we are able to observe objects up to 
a magnitude limit of $g\sim 24$, the
density of background sources will be high enough to perform
a reconstruction similar to what described in this paper.
A better approach will be to search for  peculiar fields in which 
the density is larger by chance (e.g. Petitjean 1997).
The spectral resolution will be a decreasing function of the magnitude.
Observational difficulties will include the contamination
of the LBG spectrum by absorption lines originating in the 
interstellar medium of the galaxy and the fact that the mean redshift 
($z\approx 2.8$) will be larger  than what we have considered in
this paper.
To reach these faint magnitudes we need to wait for the advent 
of the Extremely Large Telescopes (Theuns \& Srianand 2006).

To conclude, the approach developed here is very promising as the advent 
of Extremely Large Telescopes will boost this field by allowing
the observation of a  number of background sources large enough
 to probe the distribution of the matter with accurate precision
at the scales under consideration. 
The total amount of observing time will be large 
however but worthwhile given the  expected results foreseen in this paper.

%% The availability of dense enough groups of background sources required
%% to perform the reconstruction of the H~{\sc i} density field through
%% observation of Lyman-$\alpha$ absorptions is still a limitation.
%%   To reach a suitable density of the order of 300 sources per
%% square degree or more at $z>2$ requires to go to magnitudes fainter
%% than $g>23$.  This is in principle out of the reach of current
%% telescopes and instrumentation.  One possibility however would be to
%% search for a peculiar field in which the density would be larger by
%% chance (e.g. Petitjean 1997).  With the next generation of Extremely Large
%% Telescopes it will be possible to use as background sources Lyman
%% break galaxies together with quasars observed probably at different
%% spectral resolutions (Rollinde et al. 2002; Theuns \& Srianand 2006).
%% The total amount of observing time will be large however but worthwhile 
%% given the  expected results forseen in this paper.

\section*{Acknowledgements}
 We thank D. Pogosyan for providing us with his calculations of critical
 count numbers predicted in the Gaussian limit, as displayed as smooth
 curves on  Fig.~\ref{fig:criticalpts_gaussian}.  We thank  S. Prunet,
 R. Teyssier and  D. Weinberg for stimulating discussions  and
 D.~Munro  for  freely distributing   his  Yorick  programming
 language and opengl interface  (available   at  {\em\tt
 http://yorick.sourceforge.net/}).  This project was partially performed
 as a task of the HORIZON project (www.projet-horizon.fr).  The
 hydrodynamical simulations were run on the NEC-SX5 of the Institut du
 Developpement et des Ressources en Informatique Scientifique (IDRIS)
 in Orsay.

%===========
\appendix
%===========

%======================================================================
\section{Recovered field: analysis in linear space}
\label{appendix}
%======================================================================

   \begin{figure*}
     \centerline{\psfig{width={15cm},figure=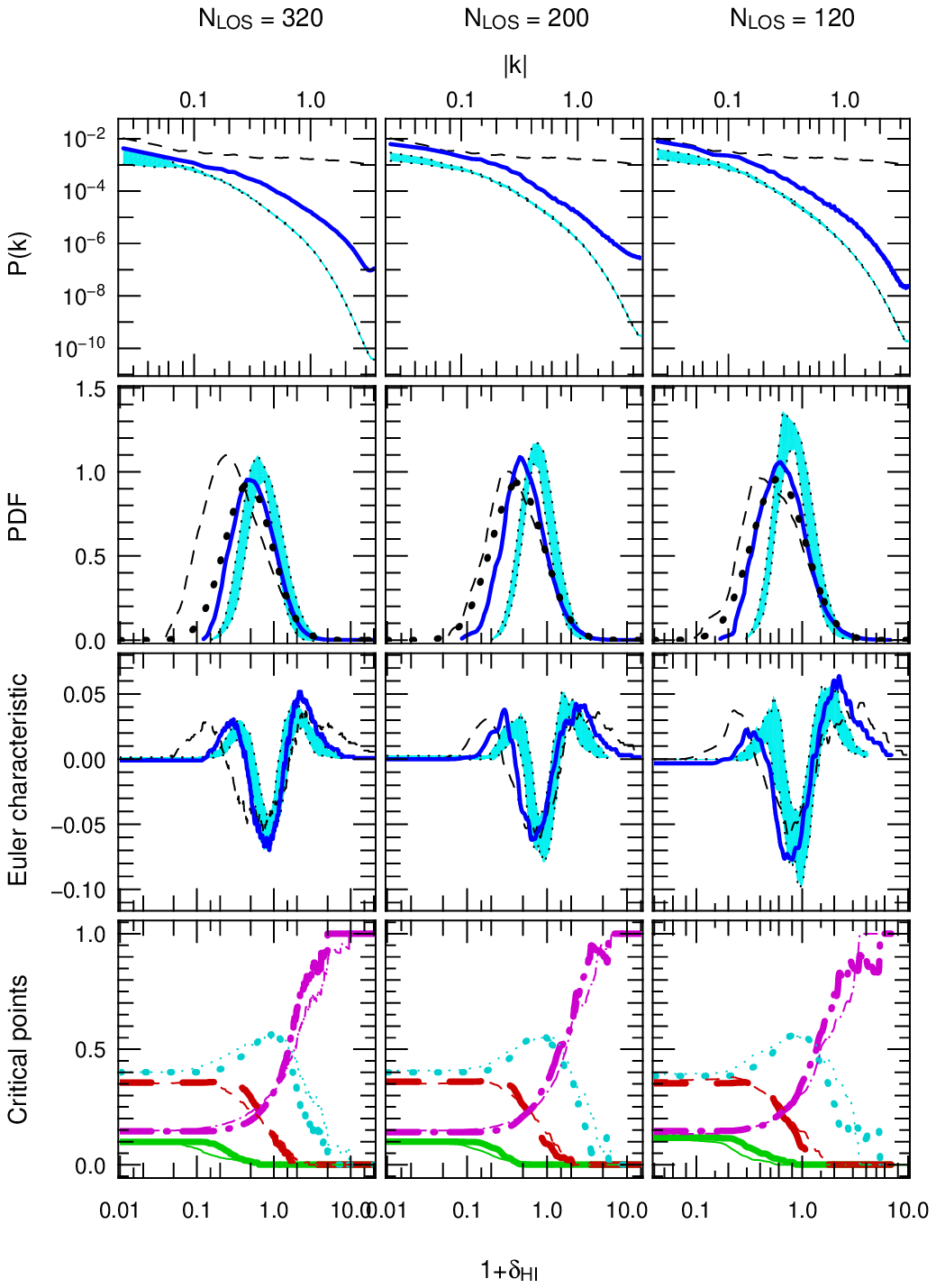}}
    \caption{Same as in Fig.~\ref{fig:power_spectrum_LOG}, but in the
    linear space, i.e. by taking the exponential of the recovered
    fields and the Gaussian realizations along with normalization
    \ref{eq:normalexp}, with subsequent smoothing with a Gaussian of
    size $\ls=\sqrt{2} \dlos$ for the last three row of panels. The
    big dots on the second row of panels now correspond to a lognormal
    distribution with same variance and average as the reconstruction.
    }
    \label{fig:power_spectrum_EXP}
   \end{figure*}

  While the reconstruction seems to perform  well for
  $\gamma=\ln(\rho)$ and its smoothed counterparts (except that it is
  somewhat  ``Gaussianized'', as shown by the measurements in the main
  text),  let us now investigate what happens for the statistical
  properties of the field  itself $\rho=\exp(\gamma)$.

  It was noted in that case (\S~\ref{sec:biasrec}) that  the recovered
  field is expected to be biased, originating from the fact that
  taking the exponential of a field does not commute with  smoothing
  via a Gaussian window. Furthermore, taking the exponential
  gives emphasis to high density peaks, which are the most poorly
  reconstructed  (\S~\ref{sec:euler_recov_fields}). Additional
  smoothing contaminates  neighboring pixels as well, resulting in
  significant changes in the connectivity. These effects were
  confirmed at the qualitative level in the main text  by visual
  inspection of Figs.~\ref{fig:slice_recorig}, \ref{fig:contour_voids}
  and \ref{fig:contour_voids_exp}. We now examine them more
  quantitatively.

\begin{figure}
  \centerline{\psfig{width={7cm},figure=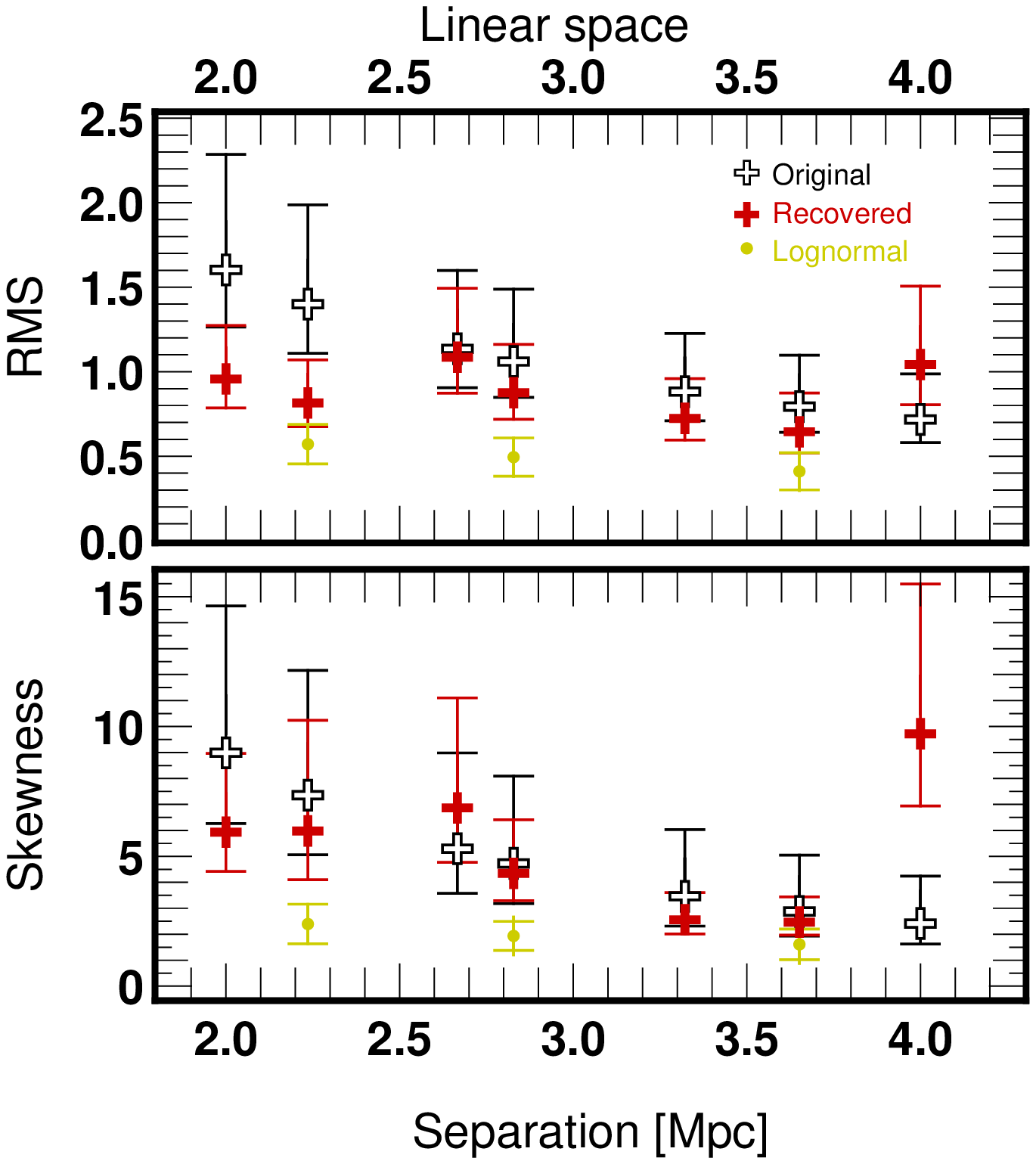}}
  \caption{Same as \Fig{skewness_rms_recorig_LOG}, but in linear
  space, as explained in caption of Fig.~\ref{fig:power_spectrum_EXP}.}
  \label{fig:skewness_rms_recorig_EXP}
\end{figure}
 \begin{figure}
    \centerline{\psfig{width={7cm},figure=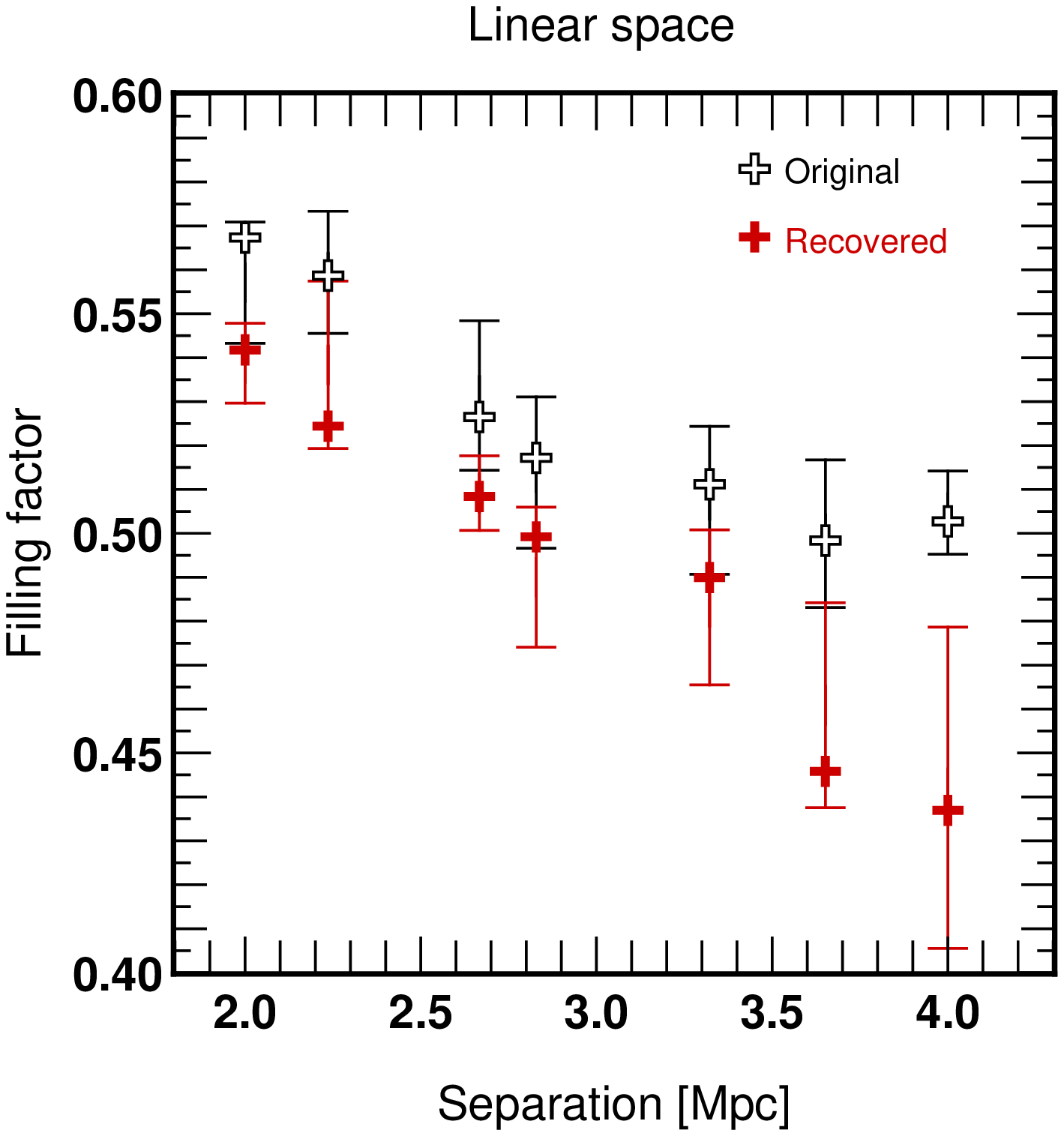}}
    \caption{Same as Fig.~\ref{fig:FFvoids_rec_orig}, but in linear space,
   as explained in caption of Fig.~\ref{fig:power_spectrum_EXP}.}
    \label{fig:FFvoids-rec-exp}
  \end{figure}
  As just argued above, since we are working in linear space, $\rho
  \sim \exp(\gamma)$, rare events in overdense regions (which are
  poorly reconstructed) dominate. As a consequence,
  the reconstruction fails with respect  to  the mean
  density: equation (\ref{eq:normrho}) is clearly not appropriate
  anymore to normalize the reconstruction. Instead, the reconstructed
  density, $\rho_{\HI,\rm rec}$, is  renormalized as follows:
  \begin{equation}
    \rho_{\HI,{\rm rec}}=\langle \rho_{\HI} \rangle \frac{\exp(
    {\tilde \gamma}_{\HI,\rm rec} )}{\langle \exp( {\tilde
    \gamma}_{\HI,\rm rec}) \rangle},
    \label{eq:normalexp}
  \end{equation}
  where $\langle \rho_{\HI} \rangle$ is the true mean density in the
  simulation. Note that this density is no longer  accurately
  determined from direct measurements on the LOSs: in the worse case
  considered in this paper, $N_{\rm LOS}=100$, we indeed find a
  relative error on the estimate of $\langle \rho_{\HI} \rangle$ of
  the order of $30$\%. However, the simulation volume is quite small,
  leading to unrealistically short LOSs.   In real observations the
  determination of the average neutral gas density along LOSs should be
  much more accurate \cite{guimaraes2007}.

  The choice of the normalization given by Eq.~(\ref{eq:normalexp}) is
  natural since it  imposes the average density of the reconstructed
  field to be equal to that of the exact solution. However, because it
  is still affected by overdense regions contributions, this
  normalization is not fully satisfactory as it does not lead to the
  appropriate corrections in underdense regions, as can be noted by
  a careful examination of 4 lower right panels of
  Fig.~\ref{fig:slice_recorig}.

  The contamination by high density peaks affects all statistics, as
  illustrated by Figs.~\ref{fig:power_spectrum_EXP} and
  \ref{fig:skewness_rms_recorig_EXP}.  This is particularly dramatic
  for second order statistics (upper row of
  Fig.~\ref{fig:power_spectrum_EXP} and upper panel of
  Fig.~\ref{fig:skewness_rms_recorig_EXP}).  The reconstruction
  underestimates the normalization of the power-spectrum, and as a
  result the variance of the PDF, especially when the separation
  between the LOSs is small: in the latter case, nonlinear features in
  the density field are given more weight and are poorly captured by
  the reconstruction.  This  appears as a shift in the PDF shown in
  the second row of panels in Fig.~\ref{fig:power_spectrum_EXP}, worsening
  with increasing $N_{\rm LOS}$. Note however that the agreement
  between the reconstruction and the exact solution,  although poorer
  than in logarithmic space, improves when $N_{\rm LOS} \la 200$. Note
  also that the smoothed lognormal fields  no longer match  the
  reconstruction.  In fact, in the linear space, its seems that the
  reconstruction gives a solution intermediate between the exact one
  and the smoothed lognormal fields, both from the point of view of
   the power-spectrum and the PDF (and its cumulants)
  (Fig.~\ref{fig:skewness_rms_recorig_EXP}):  it captures more than
  just the Gaussian features of the logarithm of the real solution, as
  would have naively followed from the analysis of
  \S~\ref{sec:euler_recov_fields}.

  These results are confirmed in the third row of panels in
  Fig.~\ref{fig:power_spectrum_EXP}: the measured Euler characteristic
  of the reconstruction gives an intermediate solution between the true
  and the lognormal solution (see for instance the position of the
  local extrema of the curves representing $\chi^{+}$). Note 
  that overall, the reconstruction matches  better the
  lognormal behaviour than the true solution, especially when $N_{\rm LOS}$
  is large,  implying that ``lognormalization'' dominates, at least
  from a topological point of view, while nonlinear dynamics implies
  significant departures from a purely lognormal behavior. This
  explains again why the quality of the reconstruction decreases when
  attempting to probe the smallest scales.  Note that this does not
  mean that decreasing the number of LOSs is better: the analysis
  always looks at the smallest scale recoverable in logarithmic space,
  $\sim 1.4 \dlos$.
  \footnote{We did not examine the skeleton in linear space to find
  the best smoothing scale in that case.}  At fixed smoothing scale, a
  reconstruction with a given number of LOSs does better than a
  reconstruction with sparser  LOS sampling.   Still, note that the
  reconstruction does more than a simple ``lognormalization'' as it
  gives an intermediary answer between the expected lognormal behavior
  from the analysis in logarithmic space and the true solution, at
  least  from the point of view of the PDF and the Euler number.  The
  uncertainties in the measurements due to the emphasis put on rare
  events are however too large to drive definite conclusions with a
  small  sample of LOS: the spread between the five lognormal fields
  is much larger than they were in the logarithmic space (and
  similarly for the PDF).

  Let us finally check the global topological properties of the
  reconstruction by examining the number counts of each kind of
  critical points individually, as shown in the last row of panels
  in \Fig{power_spectrum_EXP}.  Notwithstanding all  the above
  points,  note that the inversion achieves a  fair
  reconstruction of the distribution of some of the critical points:
  in the low density regime, it overestimates the local minima count,
  as expected from visual inspection of four lower right panels of
  Fig.~\ref{fig:slice_recorig} and from the PDF: the reconstructed
  field in underdense region is overestimated.  In the intermediate
  density range, reconstruction overestimates pancake saddle point
  counts (and to a lesser extent, underestimates filaments saddle
  point and local maxima counts) for $N_{\rm LOS}=320$ while larger
  separations between LOSs do better. In the overdense regime, where
  the reconstruction fails more dramatically, and where the amplification of
  the errors is large, one tends to overestimate (underestimate)
  filament saddle points (local maxima).

  Still, it is  interesting to note that the local minimum of the
  Euler number, $\rho_{\rm min} \sim 0.7$ is comparable for the
  reconstruction and the exact solution, suggesting that the measured
  filling factor defined previously will be similar for the  reconstruction
  and the exact solution:  according to \Fig{FFvoids-rec-exp}, the  filling
  factor of underdense regions at the minimum of the Euler number does
  nearly as well as in logarithmic space, but the match between its
  isocontours  is worse than before (compare \Fig{contour_voids} with
  \Fig{contour_voids_exp}):  thus, even if the critical point counts
  and the fraction of underdense  regions agree,
  this does not necessarily imply
  that the structures, in particular the densest ones, are at the right
  position.

\end{document}